\documentclass[aps,pre,longbibliography,showpacs,twocolumn,floatfix,superscriptaddress]{revtex4-2}

\usepackage{graphicx}
\usepackage{adjustbox}
\usepackage{dcolumn}
\usepackage{array}
\usepackage{amsmath}
\usepackage{amssymb}
\usepackage{units}
\usepackage{xcolor}
\usepackage{bm}
\usepackage{setspace}
\usepackage{color}
\usepackage{textcomp}

\newcommand{\RomanNumeralCaps}[1]

\renewcommand{\vec}[1] {\boldsymbol{#1}}
\newcommand{\pd}[2] {\frac{\partial #1}{\partial #2}}
\newcommand{\dd}[2] {\frac{\mathrm{d} #1}{\mathrm{d} #2}}
\newcommand{\DD}[2] {\frac{\mathrm{D} #1}{\mathrm{D} #2}}

\newcommand\x{\chi} 
\newcommand\X{X} 
\newcommand\rc{r} 

\DeclareMathAlphabet{\mathdutchcal}{U}{dutchcal}{m}{n}
\SetMathAlphabet{\mathdutchcal}{bold}{U}{dutchcal}{b}{n}
\DeclareMathAlphabet{\mathdutchbcal}{U}{dutchcal}{b}{n}
\DeclareSymbolFont{extraup}{U}{zavm}{m}{n}
\DeclareMathSymbol{\varheart}{\mathalpha}{extraup}{86}

\begin{document}

\title{Unmasking of Novel Conic Modes in \\
Electrically Stressed Perfectly Conducting Liquids}

\author{Chengzhe Zhou}

\affiliation{California Institute of Technology, Norman Bridge Laboratory of Physics, MC 103-33, 1200 E. California Blvd, Pasadena, CA 91125 USA}

\author{Sandra M. Troian}
\email[Corresponding author: ]{stroian@caltech.edu}
\affiliation{California Institute of Technology, T. J. Watson Sr. Laboratories of Applied Physics, MC 128-95, 1200 E. California Blvd., Pasadena, CA 91125, USA}

\homepage[]{www.troian.caltech.edu}

\date{\today}

\begin{abstract}
Liquid metal ion sources (LMIS) are widely used in applications ranging from local ion implantation in semiconductors, to focused ion beam systems for milling and nanolithography, to space micropropulsion devices being developed by NASA. Above a critically large field strength, an electrically stressed liquid metal develops one or more cuspidal protrusions which undergo accelerated conic tip sharpening with runaway field self-enhancement. Zubarev (2001) first predicted from an inviscid model that the electric stresses at the liquid apex undergo self-similar divergent growth in finite time. The inviscid assumption is appropriate to liquid metals since the viscous boundary layer extends only a few tens of nanometers from the moving interface. In this work, we examine in more depth a two-parameter family of far-field self-similar solutions incorporating inertial, electrical and capillary effects, which to leading order describe electric and velocity potential fields corresponding to a rapidly accelerating \textit{dynamic} Taylor cone. These far field solutions are incorporated self-consistently into boundary integral simulations which reveal the entire liquid shape in the near field. By invoking time reversal symmetry inherent to inviscid flow, we unmask an entire  family of novel self-similar conic modes exhibiting features such as inertial recoil, tip bulging from accelerated advance and tip counter-current flow as well as multiple interface stagnation points. These dynamic configurations help explain for the first time the origin of decades old experimental observations that have reported phenomena such as tip oscillation, pulsation and breakup during operation. The various liquid tip shapes accessible to such systems should help correct persistent misconceptions of pre- and post-emission behavior in LMIS systems and related technologies.

\end{abstract}

\maketitle

\section{Introduction}
The strong surface distortion accompanying electrically stressed liquids has fascinated researchers for centuries dating back to experiments in the early 1600's by Gilbert \cite{Gilbert1600}, who reported emission of a fine jet of liquid when water was attracted to a highly charged piece of amber, glass or thread. More than a century later, Gray \cite{Gray1731} documented similar behavior in water and quicksilver, the liquid metal now known as mercury. The development of the equations of classical electromagnetism during the 18th and 19th century by such luminaries as Lagrange, Gauss and Maxwell eventually provided the necessary framework for attempts to quantify the forces responsible for such distortion of liquid interfaces. Lord Rayleigh demonstrated in 1884 why a spherical droplet carrying a net electrical charge becomes  unstable to radial modes whenever the destabilizing Maxwell pressure due to Coulomb repulsion of surface charges exceeds the stabilizing capillary pressure. The analysis revealed that above a critical value of the surface charge or electric field, the drop is in a state of ``unstable equilibrium'' such that ultimately the  ``liquid is thrown out in fine jets''.  Seemingly unaware of Rayleigh's work, Larmor \cite{Larmor1890} in 1890 examined the behavior of capillary waves on a deep layer of electrified fluid by carrying out a linear stability analysis of the unsteady Bernoulli equation for inviscid flow. Thus he obtained a dispersion relation for the disturbance growth rate as a function of the wavelength of ripple formations on the liquid surface. His result revealed that the Coulomb repulsion of surface charges causes a reduction of the capillary wave phase velocity. These early predictions paved the way to initial quantification of the dynamics of electrified liquids but a deeper understanding of phenomena beyond the linear regime remained out of reach for decades to come.

\subsection{20th century studies by Zeleny, Larmor and Frenkel}
Advances in the century that followed began with insightful experiments by Zeleny from 1914-1920 in which he devised an elegant measurement technique based on a hydrostatic force balance \cite{Zeleny14,Zeleny15,Zeleny17,Zeleny35} to estimate the value of the electric field strength required for ejection of a fine jet. To carry out these studies, he developed a number of visualization techniques for capturing the distortion process accompanying a small hemispherical droplet protruding from the end of a fine capillary tube when exposed to a sufficiently large external field. Working with a variety of ionic liquids in ambient air and other gases and capillaries of different radii fabricated of glass and different metals, he was able to document a wide variety of phenomena including interface acceleration, retraction, oscillation and pulsation. Unlike the behavior observed in liquid metals in which emission appears to occur directly from a conic apex, Zeleny's images revealed the deformation of a small liquid mass into a very slender thread which then underwent breakup into a spray of tiny droplets. Many experimentalists using aqueous droplets and aqueous soap bubbles \cite{Nolan26,Macky30,Macky31,Nolan32,West32} soon confirmed similar behavior. In a comprehensive review of his experiments, Zeleny reported in 1935 that the intensity of the surface electric field upon discharge from either charged attached drops or uncharged drops falling in electric fields seemed to be in line with Rayleigh's original prediction for instability. However, measurement techniques at the time lacked the resolution required to determine whether the electric field strength required for instability corresponded to the value at discharge. During this period, some research groups \cite{Beams33,Quarles35} began examining in more detail the behavior of liquid mercury in vacuo to better understand differences with ionic liquids. Wanting to extract values of the field strength for ion emission from measurements of the applied voltage, these groups decided instead to use impulsive fields of the order of $10^7$ V/m applied over a very short duration period lasting anywhere from $10^{-7}-10^{-6}$ sec in an effort to prevent large scale distortion and movement of the liquid metal during testing. These studies helped catalogue estimates of the work function required for electron field emission but could not shed light on the distorted shapes just prior to emission.

By 1935, experimentalists had documented a wealth of phenomena accompanying the distortion of strongly electrified liquids. Tonks \cite{Tonks35}, a plasma physicist working at General Electric Company, was most intrigued by the fact that surface distortion and rupture leading to field emission in liquids required far smaller values of the applied field strength than for equally smooth solid surfaces. After careful re-examination of the key studies by Zeleny and others  \cite{Zeleny14,Zeleny15,Zeleny17,Nolan26,Macky30,Macky31,Nolan32,
West32,Zeleny35}, he quickly set about developing a dynamic model based on the accelerated pressure imbalance of a small hemispherical protrusion in an otherwise planar, perfectly conducting liquid subject to a critically large uniform electric field. Tonk's analysis yielded a key equation relating the amplitude of the initial liquid distortion to the so-called rupture time and field strength required, which for liquid mercury he estimated to be 53 kV/cm. His model indicated that the linear dimensions of an evolving protrusion vary inversely as the square of the field strength and that the time to rupture varies inversely as the cube of the field strength. To provide specificity, he estimated that a hemispherical bump of liquid mercury with initial maximum height $0.4~\mu\textrm{m}$ and radius $9~\mu\textrm{m}$ subject to an initial uniform field of $10^8$ V/m would sharpen in time and undergo surface rupture in about 5 microseconds.

Realizing that a rigorous analytic solution to such a complex electrohydrodynamic (EHD) problem was perhaps too formidable a task, Tonks instead relied on an insightful approximate treatment that revealed critical aspects of the distortion runaway process at late stages of development. This approach stood in contrast to all prior theoretical work, which had only investigated early time, small amplitude disturbance growth. Tonks analysis revealed that beyond a critical field strength, the pressure at the conic apex cannot support an equilibrium state since the Maxwell pressure increases as the square of the apex height. This quadratic dependence cannot therefore be counterbalanced by a stabilizing hydrostatic or capillary pressure, which at most scale linearly or inversely with apex height, respectively. This model established for the first time that any small advance of the liquid tip will continue to grow and narrow without bound until the point of emission due to the rapidly increasing unbalanced normal pressure at the conic tip. This accelerating imbalance in normal stresses causes a runaway process characterized by field self-enhancement from the ever increasing curvature at the tip. By relying on various approximations for the initial conditions and simplified functional forms for other quantities, Tonks was successful in deriving an equation for the acceleration with time of an eccentric liquid shape. Remarkably, Tonks seems also to have been the first to realize that the growth of the liquid tip should therefore proceed in self-similar fashion, as indicated by the sketches in Fig. 8 of Ref. [\onlinecite{Tonks35}], where he drew shapes depicting the accelerating interface. Shortly after Tonk's published his work, Frenkel \cite{Frenkel36} presented a slightly more general derivation based on a stability analysis of the Bernoulli equation in the inviscid limit which confirmed Tonk's equation relating the velocity of surface waves to the local Maxwell, capillary and gravitational pressure. The well-known instability describing periodic distortion of the surface of a perfectly conducting liquid above a critical field strength is now known as the Larmor-Tonks-Frenkel instability.

\subsection{Theoretical advances by Miscovsky and co-workers}
In 1964, Taylor \cite{Taylor64} set out to dispel the misconception that relations obtained from a linear stability analysis of spheroidal or other highly symmetric shapes were useful to studies of late stage deformation indicating conic like electrified shapes just prior to emission. He therefore turned attention to the polar region of a small but highly deformed liquid mass held at constant surface electric potential and sought a solution representing hydrostatic equilibrium in which the outward Maxwell pressure is everywhere exactly canceled by the inward capillary pressure. The shape corresponded to a perfect cone with interior half-angle $\theta_T \cong 49.2923^\textrm{o}$, now known as the classic Taylor angle. This result required field lines emanating from a curved counter-electrode described by the function $r/r_o = [P_{1/2}(\cos \theta)]^{-2}$, where $r_o$ is the shortest distance between the liquid apex and counter-electrode and $P_{1/2}$ denotes the Legendre function of the first kind with index $1/2$. Taylor did not discuss the fact that the solution suffered from divergence of the Maxwell and capillary pressure at the conic apex. He also did not emphasize that the majority of experimental studies, as well as the theoretical studies by Tonks and Frenkel, had established that tip sharpening is a strongly non-equilibrium and nonlinear process characterized by rapid acceleration just prior to emission. Unfortunately, this equilibrium solution led to considerable subsequent confusion in the field regarding the actual shape of dynamically evolving protrusions in electrified perfectly conducting liquids. To this day, many researchers still mistakenly believe that the tip of an electrically stressed liquid resembles a cone arising from the static balance between Maxwell and capillary pressures along the interface.

\begin{figure}[htb]
\includegraphics[scale=1]{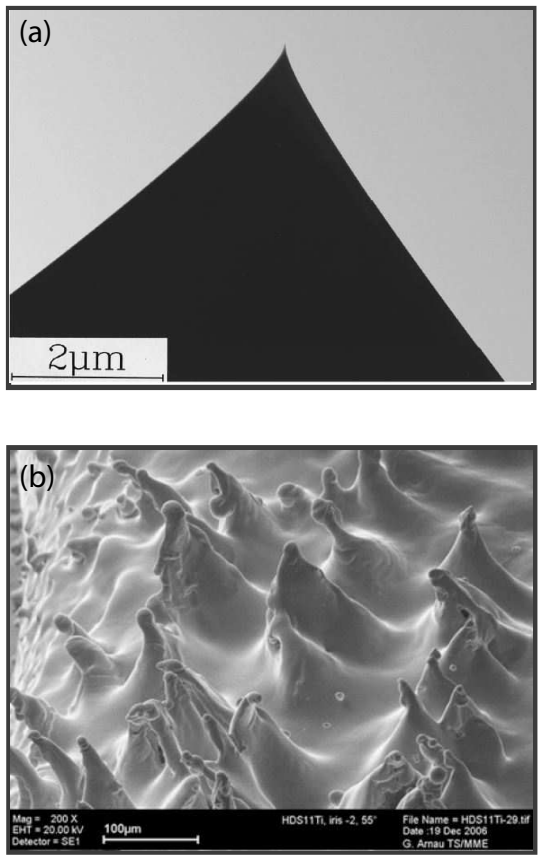}
\caption{(a) High voltage (1 MeV) transmission electron micrograph of a conic protrusion in liquid AuGe in ion emission with a 45 $\mu$A current. Reproduced with permission from Fig. 6(b) of Ref. [\onlinecite{Driesel96}]. (b) Scanning electron micrograph showing an array of protrusions in electrically stressed molten titanium imaged after solidification. The formations resulted from exposure of liquid titanium to large electric field gradients in hybrid damped RF structures within the 30 GHz Compact Linear Collider at CERN. The RF surface fields, estimated to be roughly between 95 and 135  MV/m, were applied for a duration of 70 ns every 20 ms. Reproduced with permission from Fig. 9 of Ref. [\onlinecite{Antoine12}]. \label{fg:images}}
\end{figure}	

In an effort to reconcile experimental observations with Taylor's prediction, researchers began expressing concerns over the assumptions inherent in the analysis. In 1983, Miscovsky's group \cite{Sujatha83} demonstrated that a liquid shape given by a Taylor cone is inconsistent with the equations derived from a variational formulation requiring equilibrium of perfecting conducting fluid in an electric field. Their re-examination of Taylor's derivation revealed two discrepancies with their own analysis, which included the excess fluid pressure in the normal stress boundary condition at the interface due to fluid flow and additional required terms in the Legendre expansion for the electrostatic potential beyond the first term. Using a similar variational approach \cite{Chung84}, Miscovsky and co-workers also established that even an ideal cuspidal shape is incompatible with any static equilibrium configuration. In subsequent work \cite{Chung86}, these researchers also highlighted the fact that Taylor's stability criterion represents a global and not local condition - that is, Taylor assumed that breakdown had to occur simultaneously across the entire surface of the cone, a feature at odds with most experiments showing breakdown only at a point or a series of points in extended systems. To resolve these and other problems, they set out to solve the time-dependent form of the Bernoulli equation for the interface velocity potential by incorporating inertial, Maxwell and capillary effects. This equation was then coupled to Laplace's equation for the electric potential along with the kinematic boundary condition required for mass conversation relating the surface velocity to the rate of interface displacement. This comprehensive approach gave rise to a set of nonlinear EHD equations not amenable to analytic solution in general. To make inroads, they examined stability in two distinguished limits. Their first \cite{Chung87} approach was based on the electrohydrostatic (EHS) limit in which the fields and velocities were assumed to be small to ensure a base state in quasistatic equilibrium such that the fluid surface was initially essentially at rest. The pressure difference across the interface could then be characterized by a constant value independent of position at all times. Linearization then allowed solutions describing simple, small amplitude harmonic distortion of the liquid surface from which they obtained the dispersion equation for small deformations to a base state described by a Taylor cone. Modulo the divergence at the cone apex, this analysis indicated that a static Taylor cone is an unstable configuration which should eventually disintegrate spontaneously. Their second approach was based on the EHD limit which allows for unbalanced time-dependent fluid pressure gradients to drive fluid flow. A first attempt \cite{Chung86} at a rigorous calculation valid to first order and subject to the long wavelength limit predicated a zero order shape (i.e. base state) identical to the Taylor cone solution. This calculation could proceed because a cone (and cusp) represent  coordinate surfaces of a separable coordinate system. This analysis showed that the tip of a Taylor cone should rapidly assume a concave shape with a rounded apex with a rapidly diminishing radius of curvature (i.e. precursor to a cuspidal shape) under the perturbation of an external electric field. Details of this lengthly calculation were not published until several years later when in subsequent work examining the amplification of surface capillary waves in a viscous fluid \cite{Miscovsky88,Chung89}, they extended the EHD calculation to include any wavelength disturbance. In later work, Miscovsky \textit{et al.} also examined the influence of viscous forces by carrying out a linear instability analysis of electrified thick and thin layers of fluid in an initially quiescent liquid.

Before ending this section, we wish to note an important difference in fluid configurations possible between electrically stressed ionic liquids (sometimes called leaky dielectrics) and perfectly conducting liquid metals. Recent detailed, careful simulations and experiments by two groups \cite{Burton04,Collins08,Burton11} have shed light on which systems allow and which forbid formation of progeny droplets from a conical liquid tip. Liquids with finite electrical conductivity such as ionic liquids allow for surface charge transport and redistribution, which generate surface tangential stresses leading to tip streaming. Tip streaming describes a process whereby a fluid tip elongates into a slender filament  capped by a rounded droplet which eventually detaches from the liquid thread by capillary pinchoff. Perfectly conducting or perfectly insulating liquids (like perfect dielectrics) can only sustain normal oriented electrical fields along the moving interface, which allow  formation of an accelerating self-sharpening conic tip but no tip streaming and therefore no progeny droplets. Shown in Fig. \ref{fg:images}(a) is an AuGe liquid alloy ion source resembling a conic tip observed \textit{in situ} in a 1 MeV transmission electron microscope while emitting a 45 $\mu$A current. Shown in Fig. \ref{fg:images}(b) is an scanning electron micrograph obtained after solidification of molten titanium. The numerous conic-like formations occurred after application of an extremely large field gradient of the order of 100 MV/m.

\subsection{Validity of the inviscid approximation for liquid metals}
There is strong justification for modeling the dynamic behavior of perfectly conducting liquids like viscous liquid metals using the  inviscid form of the Bernoulli equation. An initially quiescent liquid can only generate vorticity through boundary motion \cite{Batchelor00}, which then diffuses into the bulk liquid through viscous stresses initially confined within a viscous boundary layer whose thickness is about $\mu^2/\rho \sigma$, where $\mu$ denotes the liquid shear viscosity. For common liquid metals \cite{Iida88}, this viscous boundary layer is estimated to be of the order of ten nanometers. For conic formations as rapid as those observed in liquid metal ion sources, where emission occurs within a few nanoseconds to a few microseconds after startup, the viscous boundary layer is well approximated by the estimate above and therefore the majority of the bulk flow is estimated to be in the inviscid regime. Even in cases where the viscous boundary layer turns out to be much large, theoretical analysis based on the inviscid Bernoulli equation nonetheless offers valuable insight into the dynamics of complex hydrodynamic behavior. We therefore implement this approximation in our current work.

\section{Zubarev prediction of self-similar conic growth}
\label{sc:Zderivation}
In 2001, a leading physicist in the Nonlinear Dynamics Group at the Institute of Electrophysics of the Russian Academy of Sciences, developed an elegant model to probe the dynamic distortion process of an electrically stressed tip in a perfectly conducting liquid. Perhaps inspired by Tonks' \cite{Tonks35} perspective on the accelerated dynamics in that region, Zubarev \cite{Zubarev01} studied  the late time dynamics just prior to ion emission. He wondered whether the liquid apex could undergo self-similar growth culminating in a runaway process. Given the assumptions inherent to any self-similar process which require self-replicating local conditions, he required that the local electric field strength in the apical region rapidly and appreciably exceed the externally applied field strength such that only local conditions prevailed. He therefore replaced the usual external field uniformity condition by a far field boundary condition specifying vanishing field strength in the vacuum region. This insight allowed him to analyze fluid motion near the apex without reference to any particular electrode geometry, in contrast to the strong geometric constraint imposed on the counter-electrode in Taylor's original hydrostatic analysis. Scaling of the governing equations under inflation revealed a set of self-similar transformations allowing analytic expressions for the asymptotic behavior describing the  electric potential, velocity potential and interface shape. In the laboratory frame, these solutions exhibited finite time blowup in the capillary and Maxwell pressure at the liquid apex as a result of the ever diminishing radius of curvature in that region. Zubarev's approach is reviewed in detail below \cite{FN1}.

\subsection{Symbolic notation}
To keep track of notation in this current work, we here note that vector and tensor quantities are denoted by bold face variables and partial differentiation by subscripts. Dimensional variables are designated by lower case Roman or Greek letters overlay with a tilde sign (e.g. surface velocity potential $\tilde{\psi}(\tilde{\textbf{r}},\tilde{t})$), dimensionless variables by upper case Roman or Greek letters (e.g. $\Psi(\textbf{R},T)$), and dimensionless self-similar variables by lower case Roman or Greek letters (e.g. $\psi(r,z)$). The exception to these rules is the electric field distribution which is designated by $\widetilde{\vec{E}}$ in dimensional form and $\vec{E}$ in dimensionless form. Unit normal vectors along a moving interface are also assumed to be outwardly pointing from the liquid domain of interest.

Regarding the use of different axisymmetric coordinate systems to describe the dynamics of a protrusion, the original formulation of the problem as specified by Zubarev will continue to be expressed in cylindrical coordinates; however, we shall convert to a spherical coordinate system after that review. To clarify notation between coordinate systems, we have used a different font to distinguish the radial coordinate $r$ in the cylindrical system from the radial coordinate $\textsf{r}$ in the spherical system.

\subsection{Zubarev analysis based on the inviscid Bernoulli equation}
\label{sec:Zanalysis}
Here we review and expand on Zubarev's analysis for an electrically stressed axisymmetric protrusion emanating from a perfectly conducting fluid in vacuo subject to incompressible, inviscid and irrotational flow. In a perfectly conducting liquid with no net charge, all mobile charges reside on the liquid interface and rearrange instantaneously (in comparison to the timescale for fluid motion) to maintain the liquid mass at constant electric potential $\psi$. The interior liquid domain therefore defines a Gaussian volume devoid of an electric field. Consequently, the electric field at the surface of the liquid can only sustain a normal component, $\widetilde{\vec{E}}_{\tilde{n}}$ which we everywhere simply designate by $\widetilde{\vec{E}}$. Assuming all boundaries except that of the moving liquid are held stationary, the electric potential distribution in the vacuum domain varies in time only in response to the instantaneous location and shape of the moving liquid surface. (For an isolated charged liquid mass, the electric potential distribution depends only on the shape of the liquid.) The difference in the corresponding electric field distribution across the liquid/vacuum interface then gives rise to a jump in the Maxwell stress tensor $\widetilde{\vec{E}}\widetilde{\vec{E}}^{\textsf{T}} -|\widetilde{\vec{E}}|^2~\mathbb{I}/2$, which causes a jump in the electrostatic pressure \cite{Saville97} given by $- \epsilon_o \widetilde{\vec{E}}^2/2$. The negative sign reflects the fact that the net interfacial electrical stress acts to pull liquid toward the vacuum region. Setting the gauge pressure in the vacuum to be zero, the total pressure acting on the fluid interface $\tilde{p}_{\textrm{int}}$ is
\begin{equation}
\label{eq:pressure_regular}
\widetilde{p}_{\textrm{int}}(\tilde{\vec{r}},t)= -2\tilde{\mathdutchcal{H}}-\frac{1}{2}\epsilon_o |\widetilde{\vec{E}}|^2 ~,
\end{equation}
where the mean curvature in dimensional units is defined by $\tilde{\mathdutchcal{H}}= -(1/2) \widetilde{\nabla} \cdot \tilde{\vec{n}}$. We note that this expression for the interfacial pressure (\ref{eq:pressure_regular}) can also be obtained from the first variation in shape for a liquid volume whose energy is exclusively governed by electrostatic and capillary forces \cite{Ljepojevic95}.

The electric field distribution $\widetilde{\vec{E}}$ in the vacuum domain and along the moving interface $\tilde{z}=\tilde{h}(\tilde{r},\tilde{z},\tilde{t})$ is defined by the gradient of the electric potential $\tilde{\psi}(\tilde{r},\tilde{z},\tilde{t})$ satisfying Laplace's equation:
\begin{multline}
\widetilde{\nabla} \cdot \widetilde{\vec{E}}(\tilde{r},\tilde{z},\tilde{t})=0 \quad \textrm{where} ~~\widetilde{\vec{E}}=-\widetilde{\nabla}\tilde{\phi}(\tilde{r},\tilde{z},
\tilde{t})\\
\widetilde{\nabla}^2 \tilde{\phi}=\tilde{\phi}_{\tilde{r}\tilde{r}} + \frac{\tilde{\phi}_{\tilde{r}}}{\tilde{r}} + \tilde{\phi}_{\tilde{z}\tilde{z}}=0 \quad \textrm{for}~\tilde{z} \geq \tilde{h}(\tilde{r},\tilde{t})~.
\label{eqn:Laplace_phi}
\end{multline}
Likewise, for an ideal liquid subject to incompressible ($\widetilde{\nabla} \cdot \vec{\widetilde{u}} = 0$) and irrotational ($\widetilde{\nabla} \times \vec{\widetilde{u}}=0$) flow, the velocity field $\widetilde{\vec{u}}$ in the liquid domain and along the accelerating liquid interface is defined by the gradient of a velocity potential field $\tilde{\psi}(\tilde{r},\tilde{z},\tilde{t})$ satisfying Laplace's equation:
\begin{multline}
\nabla \cdot \widetilde{\vec{u}}(\tilde{r},\tilde{z},\tilde{t}) =0 \quad \textrm{where} ~~ \widetilde{\vec{u}}=(\tilde{u}, \tilde{w},\tilde{t}) =\widetilde{\nabla}  \tilde{\psi}(\tilde{r},\tilde{z},\tilde{t}) \\
\tilde{\nabla}^2 \tilde{\psi}=\tilde{\psi}_{\tilde{r}\tilde{r}} + \frac{\tilde{\psi}_{\tilde{r}}}{\tilde{r}} + \tilde{\psi}_{\tilde{z}\tilde{z}}=0 \quad \textrm{for}~~
\tilde{z} \leq \tilde{h}(\tilde{r},\tilde{t})~.
\label{eqn:Laplace_psi}
\end{multline}

Conservation of mass and momentum are enforced through the unsteady form of the inviscid Bernoulli equation, which when evaluated at the moving interface $\tilde{z}=\tilde{h}(\tilde{r},\tilde{t})$ is expressed as
\begin{align}
\rho & \left[\tilde{\psi}_{\tilde{t}} + \underbrace{\frac{1}{2}\left(\tilde{\psi}^2_{\tilde{r}} + \tilde{\psi}^2_{\tilde{z}} \right)}_\text{Inertial pressure} \right]_{\tilde{z}=\tilde{h}} = \nonumber \\
& \underbrace{\frac{\epsilon_o}{2}\left(\tilde{\phi}^2_{\tilde{r}} + \tilde{\phi}^2_{\tilde{z}}\right)_{\tilde{z}=\tilde{h}}}_\text{Maxwell pressure}+ \underbrace{\frac{\sigma}{(1+\tilde{h}_{\tilde{r}}^2)^{1/2}}
\left(\frac{\tilde{h}_{\tilde{r}\tilde{r}}}{1+\tilde{h}_{\tilde{r}}^2}+
\frac{\tilde{h}_{\tilde{r}}}{\tilde{r}}\right)}_\text{Capillary pressure} \, ,
\label{eqn:ZubarevBernoulli}
\end{align}
where $\rho$ and $\sigma$ denote the liquid density and surface tension. The underbrace terms indicate the driving pressures controlling the rate of change of the local surface velocity potential.

Zubarev introduced the following natural scalings for non-dimensionalization:
\begin{subequations}
\begin{eqnarray}
(R,Z,H) &=& \frac{\epsilon_o}{\gamma} \tilde{E}^2_o \times (\tilde{r},\tilde{z},\tilde{h}) \\
T &=& \frac{\epsilon_o^{3/2}}{\gamma\rho^{1/2}} \tilde{E}^3_o \times \tilde{t}\\
\Phi &=& \frac{\epsilon_o}{\gamma} \tilde{E}_o \times \widetilde{\phi} \\
\Psi &=& \left(\frac{\rho \epsilon_o}{\gamma^2}\right)^{1/2} \tilde{E}_o \times \widetilde{\psi}~,
\label{eqn:EndNondim}
\end{eqnarray}
\end{subequations}
where the external field $\tilde{E}_o$ was chosen to be oriented  in the vertical direction such that $\phi(z)= - \tilde{E}_o \tilde{z}$. Accordingly, the dimensionless expressions for the interface Bernoulli equation, electric potential $\Phi$ and velocity potential $\Psi$ become
\begin{gather}
\left[\Psi_T+\frac{1}{2}\left(\Psi^2_R+\Psi^2_Z\right)\right]_{Z=H} = \frac{1}{2}\left(\Psi^2_R+\Psi^2_Z \right)_{Z=H}\\
+\frac{1}{(1+H^2_R)^{1/2}}\left(\frac{H_{RR}}{1+H^2_R}+\frac{H_R}{R}\right)
\label{eqn:NDZubarevBernoulli} \\
\quad \nabla^2 \Phi (R, Z) =0 \quad Z \geq H\label{eqn:PhiLaplace}\\
\quad \nabla^2 \Psi (R, Z) =0 \quad Z \leq H\label{eqn:PsiLaplace}
\end{gather}
subject to the boundary conditions
\begin{subequations}
\begin{alignat}{3}
&\textrm{Equipotential} &~\Phi (R,Z,T)=0\quad\quad & Z = H\\
&\textrm{Decay} &\lim_{\vec{R}\to\infty} (\Phi^2_R + \Phi^2_Z)=0\quad \quad &Z>H\\
&\textrm{Decay} &\lim_{\vec{R}\to\infty} (\Psi^2_R + \Psi^2_Z)=0\quad\quad &Z<H\\
&\textrm{Symmetry} &\Phi_R (R=0,Z,T)= 0\quad \quad &Z \geq H\\
&\textrm{Symmetry} &\Psi_R (R=0,Z,T)= 0\quad \quad &Z \leq H\\
&\textrm{Symmetry} &H_R(R,T)=0\quad \quad &R=0\\
&\textrm{Kinematic} &H_T-\Psi_Z+\Psi_R H_R=0\quad\quad &Z=H.
\label{eqn:ZubarevKinematic}
\end{alignat}
\end{subequations}
Noting that the governing equations and boundary conditions remain invariant the dilations $(\Phi,\Psi)\rightarrow (\alpha \Phi,\alpha\Psi)$, $(R,Z,H)\rightarrow (\alpha^2 R,\alpha^2 Z,\alpha^2 H)$ and $T \rightarrow \alpha^3 T$, Zubarev proposed the following similarity transformations
\begin{align}
\quad \quad \quad \left[r,z,h(r,z)\right] =&~
\frac{\left[R,Z, H(R,T)\right]}{\tau^{2/3}}& \\
\quad\quad \quad \phi =&~ \frac{\Phi(R,Z,T)}{\tau^{1/3}} & z \geq h(r) \, \label{eq:sslengths}\\
\quad \quad \quad \psi =&~ \frac{\Psi(R,Z,T)}{\tau^{1/3}} &z \leq h(r) \,\\
\quad \quad \quad \textrm{where} ~~~ \tau =&~ T_\textsf{c}-T~. \quad &
\label{eqn:EndSelfSim}
\end{align}
The dimensionless blowup time $T_\textsf{c}$ defines the asymptotic time at which the apical pressures diverge to infinity due to field self-enhancement. Similar scalings \cite{Day98,Leppinen03,Sierou04} occur in inviscid model of capillary-inertial pinch-off found to  correlate well with experimental measurements of collapsing conic structures \cite{Zeff00}. The self-similar equations are given by
\begin{gather}
\left[\frac{2}{3}\!\left(r\psi_r+h\psi_z\right) -\frac{\psi}{3} +\frac{\psi^2_r+\psi^2_z}{2}\right]_{z=h} = \frac{1}{2}\left(\phi^2_r+\phi^2_z\right)_{z=h} \nonumber\\
+\frac{1}{(1+h^2_r)^{1/2}}\left(\frac{h_{rr}}{1+h^2_r}+\frac{h_r}{r}\right)
\label{eqn:NDZubarevBernoulli} \\
\quad \nabla^2 \phi (r, z) =0 \quad z \geq h\label{eqn:PhiLaplace}\\
\quad \nabla^2 \psi (r, z) =0 \quad z \leq h\label{eqn:PsiLaplace}
\end{gather}
subject to the rescaled boundary conditions
\begin{subequations}
\begin{alignat}{3}
&\textrm{Equipot.} &&\quad \phi(r,z)=0\quad && z=h\\
&\textrm{Decay} &&\quad\lim_{\vec{r}\to\infty} (\phi^2_r + \phi^2_z)=0\quad && z>h\\
&\textrm{Decay} &&\quad\lim_{\vec{r}\to\infty} (\psi^2_r + \psi^2_z)=0\quad && z<h\\
&\textrm{Symmetry} &&\quad\phi_r (r=0,z)= 0\quad && z\geq h\\
&\textrm{Symmetry} &&\quad\psi_r (r=0,z)= 0\quad\ && z\leq h\\
&\textrm{Symmetry} &&\quad h_r(r)=0\quad && r=0~~\\
&\textrm{Kinematic} &&\quad 2(rh_r-h)=3(\psi_z-h_r\psi_r)\quad &&z=h.
\label{eqn:ZubarevKinematic}
\end{alignat}
\end{subequations}
Zubarev solved these coupled equations and boundary conditions with an additional constraint, namely that the self-similar interface in the far field asymptotically conform to a conic surface given by a Taylor cone such that its exterior polar angle $\theta_0 = \pi - \theta_T$ with the classic Taylor angle $\theta_T \cong 49.2923^\textrm{o}$. In cylindrical coordinates, the asymptotic solutions for large $r$ can then be expressed as \cite{FN2}
\begin{gather}
\psi(r,z)=\sum^\infty_{n=0} a_n \frac{\partial^{3n}}{\partial z^{3n}}
\left(r^2 + z^2 \right)^{-1/2}\\
\phi(r,z)=\sum^\infty_{n=0} b_n P_{1/2}(\cos\theta)\frac{\partial^{3n}}{\partial z^{3n}}\left(r^2 + z^2 \right)^{1/4}\\
h(r)=\sum^\infty_{n=0}c_n~r^{(1-3n)}~.
\end{gather}
Here, $P_{1/2}(\cos\theta)$ is the Legendre polynomial of the first kind of order 1/2 and $\theta= \tan^{-1}(r/z)$ denotes the polar angle in spherical coordinates (and not the angular coordinate in cylindrical coordinates). In order to satisfy a vanishing equipotential on the liquid surface, it is required that $P_{1/2}(\cos\theta_o)=0$. The first few coefficients of the series are then given by
\begin{align}
&a_0 = s~\quad \quad \quad 0 < s < -\cot\theta_0\\
&b_0 =\left(\frac{dP_{1/2}(\cos\theta)}{d\theta}\right)^{-1}_{\theta_o} \left[2 (c_0-a_0) ~\right]^{1/2} \label{eq:Zb_0}\\
&c_0 =\cot\theta_0 \\
&a_1=\frac{a^2_0(1+c^2_0)^{3/2}}{18c_0(3-2c^2_0)}  \\
&b_1 =0 \\
&c_1 = 0  \\
&c_2 = \frac{a^2_0(4c^2_0 - 1)}{8c^2_0(1+c^2_0)^2(3-2c^2_0)}~.
\end{align}
For the remainder of this paper, we shall convert to a \textit{spherically symmetric} coordinate system with coordinate values $\textsf{r}, \theta)$ situated in the vacuum beyond the cone apex. When designating the value of the fluid height projected onto the axis of symmetry, it shall prove convenient to introduce the vertical projection coordinate $z=\textsf{r}\cos\theta$.

The single-valued asymptotic solutions therefore represent a \textit{one-parameter} family of solutions. The free parameter is either set by $a_0$, the leading coefficient of the velocity field, or $b_0$, the leading coefficient of the external electric field.  Accordingly,
\begin{gather}
\psi(\textsf{r},\theta) = \frac{s}{\textsf{r}}+\mathcal{O}(\frac{1}{\textsf{r}^{1/2}}) \label{eqn:sspsi}\\
\phi(\textsf{r},\theta) = b_0 ~\textsf{r}^{1/2}P_{1/2}(\cos\theta)+ \mathcal{O}(\frac{1}{\textsf{r}}) \label{eqn:ssphi}\\
h(\textsf{r}) = \textsf{r} \cot\theta_0  +\mathcal{O}(\frac{1}{\textsf{r}^5})~, \label{eqn:ssheight}
\end{gather}
where $b_0$ is given by Eq. (\ref{eq:Zb_0}). While the leading order behavior of the electrical  potential depends on both the radial and polar coordinate, the leading behavior of the velocity potential depends only on the radial coordinate. The velocity potential therefore defines a sink-like flow wherein fluid is always transported to the liquid apex in a radially symmetric manner, as depicted in Fig. \ref{fg:comparesolns}(b). The streamlines therefore always lie tangent to the liquid interface, which in the self-similar frame defines a dynamic Taylor cone with interior half-angle $\theta_T = \pi - \theta_0$. The leading correction to the perfect conic shape decays rapidly as $\textsf{r}^{-5}$. In the laboratory frame, the asymptotic shape is also given by a Taylor cone since the interior angle is invariant under the self-similar transformation defined by Eq. (\ref{eq:sslengths}); however, the spatial region near the liquid apex shrinks rapidly in time as $\tau^{2/3}$ in the limit $\tau \to 0$. The conic tip is therefore a \textit{dynamic} cone with interior flow and not a hydrostatic cone as required by Taylor's original analysis \cite{Taylor64}.

It proves useful to now estimate the magnitude of the different  contributions to the Bernoulli equation evaluated for a conic liquid shape, which can be re-expressed as
\begin{equation}
\frac{2\textsf{r}\psi_\textsf{r}-\psi}{3}+
\frac{\psi^2_\textsf{r}+\textsf{r}^{-2}\psi^2_{\theta}}{2} =
\frac{\psi^2_\textsf{r}+\textsf{r}^{-2}\psi^2_{\theta}}{2} +
\frac{\cot\theta_0}{\textsf{r}}.
\label{eqn:SSspherical}
\end{equation}
Substitution of the leading order solutions given by Eqs. (\ref{eqn:sspsi}), (\ref{eqn:ssphi}) and (\ref{eqn:ssheight}) into Eq. (\ref{eqn:SSspherical}) reveals that the second term on the left side, which defines the interface inertial pressure, scales as $\textsf{r}^{-4}$, while all the remaining terms scale as  $\textsf{r}^{-1}$. This observation indicates that to leading order, there is no kinetic contribution to the conic sharpening process, the very influence Zubarev and previous researchers intended to  incorporate into modeling efforts. Additionally to first order in $\textsf{r}$, we note that the first term on the left side, which defines the rate of change of the surface velocity potential, does not vanish. This term represents the residual flow pressure from the imbalance of the capillary and Maxwell pressure, which causes flow configurations that are not stationary in the self-similar frame. This non-vanishing term belies the original intent of seeking stationary solutions in the self-similar frame. In a subsequent publication, Zubarev \cite{Suvorov04} become aware of these two issues and resolved them successfully by proposing a different set of asymptotic solutions in which the electric and velocity potential scaled similarity as $\textsf{r}^{1/2}$. We expand further on this approach in the next section.

\section{Stationary Solutions in the Self-Similar Frame}
\label{sc:levelsetspecs}
\begin{figure}[htb]
\includegraphics[scale = 1]{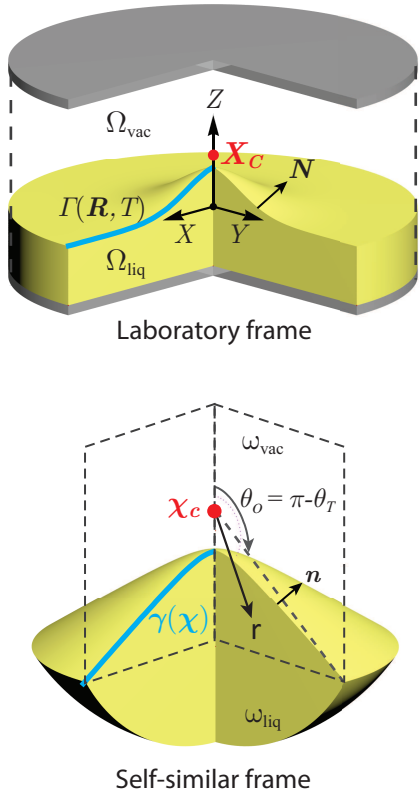}
\caption{Top image: Laboratory frame showing a snapshot of the advance of an axisymmetric protrusion with a free surface boundary  $\Gamma(\vec{R},T)$ (blue) from a pool of liquid volume $\Omega_\textrm{liq}$ in vacuo ($\Omega_\textrm{vac}$) of a perfectly conducting liquid (yellow) held at constant potential accelerating toward a circular counter electrode (grey). The (non-dimensional) Cartesian coordinates are denoted $\vec{R}=(X,Y,Z)$ - the local unit normal to the moving interface $\Gamma(\vec{R},T)$ is $\vec{N}$. The point$\vec{X_C}$ (red) indicates the blowup point at which time the capillary, Maxwell and inertial stresses undergo divergence. Bottom image: Self-similar frame for describing hydrodynamic behavior of an axisymmetric electrically stressed tip of perfectly conducting liquid (yellow) with a free boundary $\gamma(\vec{\X})$ (blue) of volume $\omega_\textrm{liq}$ advancing within a vacuum domain $\omega_\textrm{vac}$. The (non-dimensional) self-similar coordinate for a spherically symmetric system is denoted $\vec{\X}=(\textsf{r},\theta)$ - the local unit normal to the free surface boundary is $\vec{n}$. The blowup point is $\vec{\X}_c$. The polar angle corresponding to a conic envelope with Taylor angle $\theta_T \cong 49.2923^o$ is given by $\theta_0 = \pi - \theta_T$ where the Taylor angle.
\label{fg:LBvsSS}}
\end{figure}

In what follows, we derive the asymptotic behavior of self-similar solutions characterizing the sharpening tip of an electrically stressed liquid, generalized to include inertial effects and time reversal symmetry. Shown in Fig. \ref{fg:LBvsSS} is the geometry and coordinate system of choice in referring to the (dimensionless) laboratory frame and (dimensionless) self-similar frame described by  spherically symmetric coordinates $(\textsf{r},\theta)$. The free surface in the laboratory frame is defined by a Cartesian based level set function $F(\vec{\X}, T)$ where the zero set of $F(\vec{\X}, T)= Z - \Gamma(X,Y,T)$ defines the time-dependent interface function $\Gamma$ given by:
\begin{equation}
\label{eq:GammaLevelSet}
\Gamma =\left \{\vec{\X} | F(\vec{\X}, T) = 0\right\}~.
\end{equation}
From the scalings in Eqs. (\ref{eqn:EndNondim}), the non-dimensional Bernoulli equation for the velocity potential $\Psi$, valid throughout the interior of the liquid domain $\Omega_{\textrm{liq}}$ and on the liquid interface $\Gamma$ is given by
\begin{equation}
\label{eq:bernoulli_bulk}
\pd{\Psi}{T}+\frac{1}{2}\nabla\Psi\cdot \nabla\Psi+P = 0~,
\end{equation}
where $P$ denotes the fluid pressure, which when evaluated at the moving interface, is expressed by
\begin{equation}
\label{eq:pressure_regular} P=-2\mathcal{H}-\frac{1}{2}|\vec{E}|^2\quad\textrm{on}\quad\Gamma~.
\end{equation}
The kinematic boundary condition in Eq. (\ref{eqn:ZubarevKinematic}) requires that the material points on the free surface move according to the local value of the normal component of the liquid surface velocity. For the level set representation given by Eq. (\ref{eq:GammaLevelSet}), the function $F(\vec{X},T)$ is therefore advected by the surface velocity field $\vec{U}_{\Gamma}$ according to
\begin{equation}\label{eq:kinematic_regular}
\DD{F}{T}=0\quad\textrm{on}\quad\Gamma~,
\end{equation}
where $\mathrm{D}/\mathrm{D}T = \partial /\partial T + \vec{U}_{\Gamma} \cdot \nabla$ denotes the material derivative. All other equations and boundary conditions specified in Section \ref{sc:Zderivation} can be similarly non-dimensionalized.

Given the assumption of irrotational inviscid flow, we generalize the analysis to allow for time reversal symmetry by introducing the dimensionless time scale $\tau$ defining pre- and post-singularity flow:
\begin{align}
\tau & = +~(T_C-T)~\textrm{for}~ T< T_C ~\textrm{pre-singularity flow}\\
\tau & = -~(T_C-T) ~ \textrm{for}~ T> T_C ~\textrm{post-singularity flow}
\end{align}
where the apical singularity occurs at the blowup time $T_C$. For $\tau=+(T_C-T)$, the fluid interface advances toward the singularity  from below, which we coin a pre-singularity event. For  $\tau=-(T_C-T)$, the fluid interface is retracting toward the singularity from above - a post-singularity event. From time-reversal symmetry, advancement toward the blowup point is equivalent to retraction toward the blowup point. Additional discussion appears in Section \ref{sec:dynselfsim}.

As before, the self-similar coordinate vector $\vec{\x}$ which preserves dilational symmetry is defined by
\begin{equation}
\label{eq:tau_eta}
\vec{\x} = \frac{\vec{\X}_c - \vec{\X}}{\tau^{2/3}}~.
\end{equation}
The semi-infinite liquid domain $\omega_\mathrm{liq}$ is separated from the semi-infinite vacuum domain $\omega_\mathrm{vac}$ by the level set function $\gamma(\vec{\X})$ where
\begin{gather}
\label{eq:F_Phi_Psi}
\psi(\vec{\x},t) = \pm \frac{\Psi(\vec{\X},T)}{\tau^{1/3}} \\
\phi(\vec{\x},t) = \frac{\Phi(\vec{\X},T)}{\tau^{1/3}} \\
f(\vec{\x},t) = \frac{F(\vec{\X},T)}{\tau^{2/3}} ~.
\end{gather}
As $\tau \to 0$, the velocity and electric field potential increase rapidly while all length scales including the function $\gamma = \left\{\vec{\x} \mid f(\vec{\x},t) = 0\right\}$ and the apex curvature radius (not shown) decrease rapidly, reflecting tip sharpening. The general scalings here allow for additional time dependence in $\psi$, $\phi$ and $f$ over and above algebraic growth in $\tau$. Because of the isotropic rescaling of spatial coordinates, the Laplace equation for the velocity and electric potential remain unchanged:
\begin{equation}
\label{eq:potential_selfsimilar}
\nabla^2\psi = 0 \quad\textrm{in}\quad\omega_\mathrm{liq} ~~~~\textrm{and}~~~~
\nabla^2\phi = 0 \quad\textrm{in}\quad\omega_\mathrm{vac}~.
\end{equation}
The differential operator $\nabla$ is understood to act on the self-similar coordinate $\vec{\x}$. Transformation to the self-similar frame yields the rescaled Bernoulli, interface and equipotential equations:
\begin{gather}
\pd{\psi}{t}\!+\!\frac{2}{3}\vec{\x}\cdot \nabla\psi-\frac{\psi}{3}+\frac{1}{2}\left|\nabla \psi\right|^2 = 2\mathdutchcal{h}\!+\!\frac{1}{2}\left|\nabla \phi\right|^2 \quad\textrm{on}~\gamma~,
\label{eq:unsteadyBernoulliSS} \\
\frac{1}{|\nabla f|}\pd{f}{t}+\frac{2}{3}\vec{n}\cdot\vec{\x}+\vec{n}\cdot\nabla\psi =0\quad\textrm{on}~\gamma~
\label{eq:unsteadyKinematicSS} \\
\phi =\textrm{constant}	\quad\textrm{on}~\gamma~.
\label{eq:unsteadyEquipotentialSS}
\end{gather}
We introduce the dilated time variable $t(T) = - \ln \tau$ to allow stretching or slowdown of the algebraically fast dynamics. The quantity $\mathdutchcal{h}$ denotes the mean curvature of the boundary $\gamma$ and $\vec{n} = \nabla f /|\nabla f|$ denotes the outwardly pointing unit normal vector along $\gamma$. The hydrodynamic quantities of interest in the laboratory frame are recovered from the relations
\begin{align}	 \vec{U}(\vec{\X},T)&=\pm\frac{\nabla\psi(\vec{\x},t)}{\tau^{1/3}}\\
\vec{E}(\vec{\X},T)&=- \frac{\nabla\phi(\vec{\x},t)}{\tau^{1/3}}\\
P(\vec{\X},T)&=\frac{p(\vec{\x},t)}{\tau^{2/3}}~.
\label{eq:UE}
\end{align}
From Eq. (\ref{eq:bernoulli_bulk}), the pressure within the liquid is given by
\begin{equation}
p(\vec{\x},t)=-\pd{\psi}{t}-\frac{2}{3}\vec{\x} \cdot \nabla \psi +\frac{\psi}{3} -\frac{1}{2}\lvert\nabla\psi\rvert^2~.
\label{eq:pressure_selfsimilar}
\end{equation}
We restrict our analysis to flow configurations that are  \textit{stationary} in the self-similar frame. In spherically symmetric coordinates, the interface Bernoulli equation given by
\begin{equation}
\frac{2\textsf{r}\psi_\textsf{r}\!-\psi}{3}\!+\!
\frac{\psi^2_\textsf{r}\!+\!\textsf{r}^{-2}\psi^2_{\theta}}{2}\!=\!2~\mathdutchcal{h}\!+\! \frac{\phi^2_\textsf{r}\!+\!\textsf{r}^{-2}\phi^2_{\theta}}{2} \quad\quad \textrm{on}~\Theta(\textsf{r})
\label{eqn:SphericBern}
\end{equation}
couples to the harmonic equations
\begin{align}
&\phi_{\textsf{rr}}+\frac{2\phi_\textsf{r}}{\textsf{r}}+
\frac{\phi_{\theta\theta}}{\textsf{r}^2}+\frac{\cot\theta \, \phi_{\theta}}{\textsf{r}^2}=0 \quad & \theta \geq \Theta(\textsf{r})
\label{eqn:SphericPhiLaplace}\\
&\psi_{\textsf{rr}}+\frac{2\psi_\textsf{r}}{\textsf{r}}+
\frac{\psi_{\theta\theta}}{\textsf{r}^2}+\frac{\cot\theta \,\psi_{\theta}}{\textsf{r}^2}=0 ~\quad & \theta \leq \Theta(\textsf{r}),
\label{eqn:SphericPsiLaplace}
\end{align}
where $\Theta(\textsf{r})$ defines the free boundary separating the liquid from vacuum domain. The capillary pressure $2~\mathdutchcal{h} = - \nabla \cdot \vec{n}$ is given by
\begin{align}
\nabla \cdot \vec{n} &= \frac{\partial n_\textsf{r}}{\partial \textsf{r}}+\frac{2n_\textsf{r}}{\textsf{r}} + \frac{n_\theta \cot\theta}{\textsf{r}}\quad \textrm{where} \\
n_\textsf{r} &= \frac{-~\Theta(\textsf{r})}{(\Theta^2_\textsf{r} + \textsf{r}^{-2})^{1/2}}\\
n_\theta &= \frac{1}{\textsf{r}(\Theta^2_\textsf{r} + \textsf{r}^{-2})^{1/2}}~.
\end{align}
These equations are subject to the following boundary conditions
\begin{subequations}
\begin{alignat}{4}
&\textrm{Equipot.}&&\quad\ \phi(\textsf{r},\theta)=0\quad\quad\quad\quad\quad\quad ~\textrm{on}~\Theta(\textsf{r})&&\label{eq:SSequipot}\\
&\textrm{Decay}&&\quad\lim_{\textsf{r}\to\infty} \left(\phi^2_\textsf{r}+\frac{\phi^2_{\theta}}{\textsf{r}^2}\right)=0\quad\quad \theta > \Theta(\textsf{r})&&\\
&\textrm{Decay}&&\quad \lim_{\textsf{r}\to\infty}\left(\psi^2_\textsf{r}+\frac{\psi^2_{\theta}}{\textsf{r}^2}\right)=0 \quad\quad\theta <\Theta(\textsf{r})&&\\
&\textrm{Symmetry}&&\quad\quad\quad\phi_{\theta}(\textsf{r},\theta)=0\quad\quad\quad\quad\quad\theta=0,\pi&&\\
&\textrm{Symmetry}&&\quad\quad\quad\psi_{\theta}(\textsf{r},\theta)=0\quad\quad\quad\quad\quad\theta=0,\pi&&\\
&\textrm{Symmetry}&&\quad\quad\quad\Theta_\textsf{r}(\textsf{r})=0\quad\quad\quad\quad\quad\quad\theta=0&&\\
\label{eq:SSkin}
&\textrm{Kinematic}&&\!\!\quad\frac{2}{3}\Theta_\textsf{r}\!=\! \frac{\psi_{\theta}}{\textsf{r}^3\sin^2\Theta}\!-\!
\frac{\psi_{\theta}\Theta_\textsf{r}\cot\Theta}{\textsf{r}^2}\!-\!\frac{\psi_\textsf{r} \Theta_\textsf{r}}{\textsf{r}}.\quad &&&
\end{alignat}
\end{subequations} As $\textsf{r} \to \infty$, Eq. (\ref{eq:SSkin}) reduces simply to  $\Theta_r=0$, which defines a perfect conic surface with constant exterior polar angle $\theta_0$. We also note from Eq. (\ref{eqn:SphericBern}) that the term $(2/3) \vec{\x}\cdot\nabla\psi$ reduces simply to $(2/3)\textsf{r}\psi_\textsf{r}$ since the polar angle $\theta$ is invariant under the self-similar transformation.

\subsection{General Features of Asymptotic Potential Fields in Three Special Limits}
\label{sc:famAsymDynCone}
The self-similar equations and boundary conditions specified are used to elucidate three limiting behaviors, depending on the form of the velocity potential and whether inertial effects are incorporated, as discussed next. One such limit is Zubarev's original solution described in Section \ref{sec:Zanalysis}, given by Eqs. (\ref{eqn:sspsi}) - (\ref{eqn:ssheight}), which are depicted in Fig. \ref{fg:comparesolns}. For the purpose here, the most important characteristic of this solution is the lack of contribution from inertial effects, which restricts the configurations possible for the advancing fluid tip.

\subsection{Classic hydrostatic Taylor cone solution}
\label{sec:taylorlike}
The classic hydrostatic Taylor cone solution describes a stationary electrified fluid in the laboratory frame which is acted upon solely by capillary and Maxwell forces. Consider the limit in which the velocity potential is everywhere constant throughout $\omega_{liq}$ and on $\gamma$, which without loss of generality can be specified to be $\psi = 0$. The general harmonic solution \cite {Jackson99} for the electric potential $\phi$ is known to be
\begin{equation}
\label{eq:psi0}
\phi(\textsf{r},\theta)=\sum^\infty_{\nu=0}b_n \textsf{r}^\nu P_\nu(\cos\theta)~,
\end{equation}
where $P_\nu(\cos\theta)$ is the Legendre function of the first kind of order $\nu$, $P_\nu(\cos\theta)=P_{-\nu-1}(\cos\theta)$ and $\nu$ denotes a real non-integral value. The functions $P_\nu(\cos\theta)$ have a logarithmic singularity at $\theta=\pi$, which is excluded since $\phi$ is confined to the vacuum domain $0 \leq \theta \leq \Theta(\textsf{r})$.  For $\phi$ to be finite at the origin, $\nu$ must be positive. The requirement that the liquid represent an equipotential mass requires $P_\nu(\cos\theta)=0$. Substitution of $\psi=0$ (as well as its derivatives) into Eq. (\ref{eqn:SphericBern}) reduces the Bernoulli equation to the two competing terms on the right hand side reflecting the balance required between the capillary and Maxwell pressure. Since the mean curvature $\mathdutchcal{h}$ scales as $1/\textsf{r}$ and the Maxwell pressure scales as $|\nabla\psi|^2\sim \textsf{r}^{2(\nu-1)}$, the only allowable solution is $\nu=1/2$. For $P_1/2$, the only zero in the range $0 < \theta < \pi$ is given by $\cos \theta_0 = -0.6522$ so that $\theta_0 \cong 130.7077$ and $\pi - \theta_0 \cong 49.2923^\circ = \theta_T$. To leading order, this stationary solution is then given by
\begin{align}
\psi=& ~0 & \theta \leq \theta_0 \label{eq:taylorSolutionpsi}\\
\phi=& ~b_0~\textsf{r}^{1/2} P_{1/2}(\cos\theta) & \theta \geq \theta_0, \label{eq:taylorSolutionphi}
\end{align}
where from Eq. (\ref{eqn:SphericBern}) the constant $b_0$ is evaluated to be
\begin{equation}
\label{eq:b0}
b_0=\frac{\sqrt{-2\cot\theta_0}}{\mathrm{d}P_{1/2}(\cos\theta)/
\mathrm{d}\theta|_{\theta=\theta_0}}\approx 1.34593~.
\end{equation}
The solution, plotted in Fig. \ref{fg:comparesolns}, represents the original Taylor solution \cite{Taylor64}. Note that it contains no free parameters. More importantly, however, the solution is not uniformly valid since the electric field strength $ \partial \phi/ \partial textsf{r}$ diverges to infinity as $\textsf{r} \to 0$.

\subsection{General self-similar solution \\ with far field conic shape}
\label{sec:dynselfsim}
Stationary solutions in the self-similar frame allow nonetheless allows liquid configurations in the laboratory frame capable of rapid acceleration. For this reason, the most general asymptotic solutions  should enable that inertial, capillary and Maxwell pressures all contribute to leading order in Eq. (\ref{eqn:SphericBern}).
\begin{figure}[htb]
\includegraphics[scale=1]{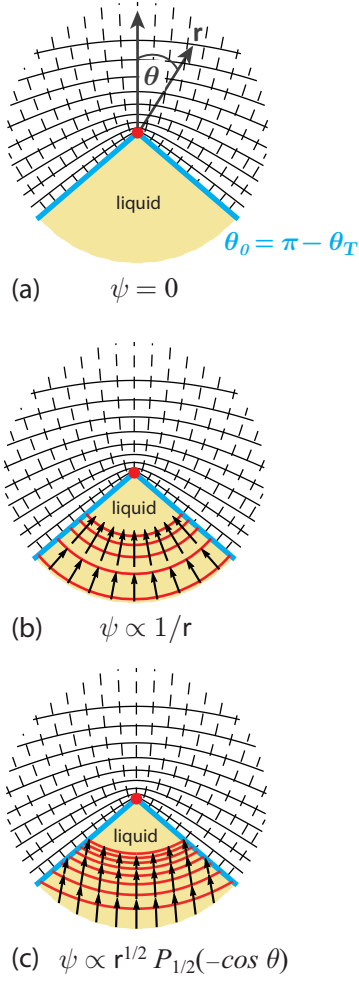}
\caption{Asymptotic solutions depicting the electric potential field  ($\psi(\textsf{r},\theta$) (solid black lines), electric force field  (dashed black lines), fluid pressure p (solid red lines) and fluid velocity $\nabla \psi$ (black arrows). Where shown, the isobaric contours (red) are separated by equal pressure increments.The liquid domain (yellow) is bounded above by the vacuum/liquid interface (blue line) which in the far field assumes the form $h = \cot\theta_0~\textsf{r}$. The conic apex point (red) represents the intersection of the central symmetry axis with the line tangent to the cone envelope. (a) Classic Taylor solution characterized by a vanishing velocity field. (b) Self-similar sink flow solution given by Eqs. (\ref{eqn:sspsi}), (\ref{eqn:ssphi}) and (\ref{eqn:ssheight}) describing purely radial conic flow. (c) General self-similar solution given by Eqs. (\ref{eqn:psiSS}), (\ref{eqn:phiSS}) and (\ref{eq:ansatz}) describing fully radial and angular dependent flow confined to a conic domain.}
\label{fg:comparesolns}
\end{figure}
Inspection of Eq. (\ref{eqn:SphericBern}) reveals this is possible as $\textsf{r }\to \infty$ if $\phi \sim \textsf{r}^{1/2} F(\theta)$ and $\psi \sim \textsf{r}^{1/2} G(\theta)$ where $F$ and $G$ are spherically symmetric Legendre functions of order 1/2. Substitution then leads to cancelation of the first term on the left side since $2\textsf{r}\psi_\textsf{r} = \psi$, leaving only the terms proportional to $\textsf{r}$, namely
\begin{equation}
\frac{1}{r}\left( \frac{G^2}{8} + \frac{G^2_\theta}{2} \right) = \frac{1}{r}\left( \frac{F^2_\theta}{2} + \cot\theta_0 \right)~.
\end{equation}
The scaling $\psi \sim \textsf{r}^{1/2} G(\theta)$ has been used for studies of capillary pinch-off \cite{Day98} and dynamic cone formation including inertial effects  \cite{Suvorov04}. The solutions to the Bernoulli equation to leading order are given by
\begin{align}
\psi(\textsf{r},\theta)= & ~ a_0~ \textsf{r}^{1/2}P_{1/2}(-\cos\theta)
+\frac{a_1}{\textsf{r}} + \mathcal{O}(\frac{1}{\textsf{r}^{5/2}}) \label{eqn:psiSS}\\
\phi(\textsf{r},\theta)= & ~ b_0~ \textsf{r}^{1/2} P_{1/2}(\cos\theta) +\mathcal{O}(\frac{1}{\textsf{r}}) \label{eqn:phiSS}\\
\Theta(\textsf{r})= & ~\theta_0~ + \mathcal{O}(\frac{1}{\textsf{r}^{3/2}})\\
h(\textsf{r},\theta)= & \textsf{r}\cos\Theta(\textsf{r}) = \textsf{r}\cos\theta_0 + \mathcal{O}(\frac{1}{\textsf{r}^{1/2}})~.
\label{eq:ansatz}
\end{align}
The series expansions proportional to $Q_{1/2}$ (Legendre function of the second kind of order 1/2) are eliminated to prevent divergence of $\phi$ at $\theta=0$ and $\psi$ at $\theta=\pi$. The term $P_{1/2}(-\cos\theta)$, which has a logarithmic singularity at $\theta=0$, is non-divergent since the velocity potential is confined to the liquid domain $0 < \Theta(\textsf{r}) \leq \theta \leq \pi$. Likewise, the term $P_{1/2}(\cos\theta)$, which has a logarithmic singularity at $\theta=\pi$, is non-divergent since the electric potential is confined to the vacuum domain $0 \leq \theta \leq \Theta(\textsf{r})$. Substitution of these expansions into the kinematic equation in Eq. (\ref{eq:SSkin}) yields the useful relation
\begin{equation}
\label{eq:a0_a1_b0}
\begin{split}
a_1&=\frac{a_0^2}{2}\bigg\{\bigg[\frac{P_{1/2}(-\cos\theta_0)}{2}\bigg]^2 +\bigg[\dd{P_{1/2}(-\cos\theta)}{\theta}\bigg|_{\theta_0}\bigg]^2\bigg\} \\
&\quad-\frac{b_0^2}{2}\bigg[\dd{P_{1/2}(\cos\theta)}
{\theta}\bigg|_{\theta_0}\bigg]^2-\cot\theta_0~.
\end{split}
\end{equation}
Shown in Fig. \ref{fg:comparesolns}(c) is the velocity potential $\psi \propto r^{1/2}~P_{1/2} (- \cos \theta)$. Due to inertial effects, the field lines are no longer oriented radially nor aligned with the pressure gradient. Equations (\ref{eqn:psiSS}), (\ref{eqn:phiSS}) and (\ref{eq:ansatz}) specify only the leading order behavior. The full series expansions are given by
\begin{gather}
\label{eq:psiexpansion}
\psi_\infty(\textsf{r},\theta) = \sum^\infty_{k=0}a_k~ \textsf{r}^{\frac{1}{2}-\frac{3}{2}k}~P_{\frac{3}{2}k-\frac{3}{2}}(-\cos\theta) \\
\label{eq:phiexpansion}
\phi_\infty(\textsf{r},\theta) = \sum^\infty_{k=0}b_k ~ \textsf{r}^{\frac{1}{2}-\frac{3}{2}k}~P_{\frac{3}{2}k-\frac{3}{2}}(\cos\theta) \\
\label{eq:hexpansion}
h_\infty(\textsf{r}) = \sum^\infty_{k=0} c_k~\textsf{r}^{1-\frac{3}{2}k}~.
\end{gather}
As shown below, the  general recursion relations for $a_k$, $b_k$ and $c_k$ for $k > 1$ are uniquely determined by the leading order coefficients $a_0$, $b_0$ and $c_0$. However, since the asymptotic slope of the liquid mass $c_0$ must equal $\theta_T$, the coefficient relations are generated by only two independent parameters. The procedure for determining higher order coefficients is a nontrivial exercise since $a_k$, $b_k$ and $c_k$ are coupled together by three nonlinear equations which must be evaluated on the exact interface shape $h=c_0 \textsf{r}+\sum_{k=1}^{\infty}c_k h_k$ -- and not the simple Taylor cone shape $h=\textsf{r}\cos\theta_0$. The interweaved procedure for obtaining these coefficients in illustrated in Fig. \ref{fg:abcFlowChart}. These coefficients were computed term by term using a symbolic manipulation software package \cite{Mathematica19}. The coefficients for the velocity potential are
\begin{widetext}
\begin{equation}
\def\localIndent{\mathbin{\phantom{=}}}
\left.\begin{aligned}
a_0&=\,\textrm{Free parameter}\\
a_1(a_0,b_0,c_0)&= \,\textrm{Eq.} (\ref{eq:a0_a1_b0})\\
a_2(c_0,\dots,c_3,a_1)&=\,\frac{3 }{\sqrt{\sin \theta_0} P_{3/2}'(-\cos \theta_0)}
\left(-\frac{a_1 c_1}{2  }
+\frac{c_1^3}{8 \sin \theta_0 }
+\frac{c_3}{\sin ^{3}\theta_0 }
\right)~.
\end{aligned}
\label{eq:a0a1a2}
\right\}
\end{equation}
\end{widetext}
The coefficients for the electric potential are
\begin{widetext}
\begin{equation}
\def\localIndent{\mathbin{\phantom{=}}}
\left.\begin{aligned}
b_0&=\,\textrm{Free parameter},\\
b_1(c_0,c_1,b_0)&=-b_0 c_1 \sin ^{{3}/{2}}\theta_0 P_{1/{2}}'(\cos \theta_0),\\
b_2(c_0,\dots,c_2,b_0)&=-b_0
\frac{ 2 c_2 P_{1/2}'(\cos \theta_0)+c_1^2 \sin ^3\theta_0 P_{1/2}''(\cos \theta_0)}{2 P_{3/2}(\cos \theta_0)},\\
b_3(c_0,\cdots,c_3,b_0)
&=b_0 P_{1/2}'(\cos \theta_0) \left[\frac{c_1 c_2 P_{3/2}'(\cos \theta_0)}{\csc ^{ {3}/{2}}\theta_0 P_{3/2}(\cos \theta_0) P_3(\cos \theta_0)}\right.\\
&\localIndent\left.-\frac{ 8 c_3 \csc ^4\theta_0+c_1^3 \left(\csc ^2\theta_0+1\right)+12 c_1 c_2 \cot \theta_0 \csc ^2\theta_0}{8 \csc ^{ {5}/{2}}\theta_0 P_3(\cos \theta_0)}\right]\\
&\localIndent+~b_0 P_{1/2}''(\cos \theta_0)
\left[\frac{ c_1^3 P_{3/2}'(\cos \theta_0)}{2 \csc ^{ {9}/{2}}\theta_0 P_{3/2}(\cos \theta_0) P_3(\cos \theta_0)}\right.\\
&\localIndent\left.~-\frac{c_1 c_2}{\csc ^{ {3}/{2}}\theta_0 P_3(\cos \theta_0)}\right]\\
&\localIndent-~b_0P_{1/2}^{'''}(\cos \theta_0)\frac{c_1^3 }{6 \csc ^{ {9}/{2}}\theta_0 P_3(\cos \theta_0)}.
\end{aligned}
\label{eq:b0b1b2}
\right\}
\end{equation}
\end{widetext}
The coefficients for the liquid interface height are
\begin{widetext}
\begin{equation}
\def\localIndent{\mathbin{\phantom{=}}}
\left.\begin{aligned}
c_0&=\, \cos{\theta_0}\\
c_1(a_0,c_0)&=\frac{P_{-3/2}'(-\cos \theta_0)}{ \sqrt{\csc \theta_0}}  a_0,\\
c_2(c_0,c_1)&=-\frac{c_1^2}{2 P_{1/2}'(-\cos \theta_0)/\sin \theta_0}
\left[ 2 \cos ( \theta _0) P_{1/2}'(-\cos \theta_0)\right.\\
&\left.\localIndent +\sin ^2\theta_0 P_{1/2}''(-\cos \theta_0)+3P_{1/2}(-\cos \theta_0)/4\right].
\end{aligned}
\label{eq:c0c1c2}
\right\}
\end{equation}
\end{widetext}
In the recursion relations shown below, the notation $P'_{\nu}(\cdot)$ refers to differentiation of the Legendre function with respect to the argument $\cos\theta$. Equation (\ref{eq:c0c1c2}) shows that $c_1$ and $a_0$ are simply related by a constant. The general self-similar solution therefore specifies a \textit{two-parameter } family parameterized either by $(a_0,b_0)$ or equivalently $(c_1, b_0)$. For example then, the coefficient $a_2$ can be expressed in terms of $b_0$ and $c_1$ where
\begin{widetext}
\begin{align}
\notag&a_2(c_1,b_0)=c_1 \left[b_0^2
\frac{P_{1/2}(-\cos \theta_0)-4 \cos \theta_0 P_{1/2}'(-\cos \theta_0)}{8 \csc ^{3/2}\theta_0 P_{3/2}(-\cos \theta_0) P_{1/2}'(-\cos \theta_0)} P_{1/2}'\left(\cos \theta_0\right)^2\right.\\
\notag&\left.+\frac{ 4 \cos \theta_0 P_{1/2}(-\cos \theta_0)-\left(3 \cos \left(2 \theta _0\right)+5\right) P_{1/2}'(-\cos \theta_0)}{16 P_{3/2}(-\cos \theta_0) P_{1/2}'(-\cos \theta_0)}\csc ^{3/2}\theta_0\right]\\
&+c_1^3\frac{ 16 \cos \theta_0 P_{1/2}'(-\cos \theta_0)^3+4  P_{1/2}(-\cos \theta_0) P_{1/2}'(-\cos \theta_0)^2-\csc ^2\theta_0 P_{1/2}(-\cos \theta_0)^3}
{32 \csc ^{1/2}\theta_0 P_{3/2}(-\cos \theta_0) P_{1/2}'(-\cos \theta_0)^3}.
\end{align}
\end{widetext}
A simple change of variable $-\cos \theta_0\to x$ in Eq. (\ref{eq:c0c1c2}) also reveals that $c_2=0$ since
\begin{align}
c_2&\propto2 \cos ( \theta _0) P_{1/2}'(-\cos \theta_0)+\sin ^2\theta_0 P_{1/2}''(-\cos \theta_0) \nonumber \\
&+\frac{3}{4}P_{1/2}(-\cos \theta_0) \nonumber \\
&=\dd{}{x}\left[(1-x^2)\dd{P_{1/2}(x)}{x}\right]+\frac{1}{2}
\left(\frac{1}{2}+1\right)P_{1/2}(x)
&=0.
\end{align}
The last equality derives from the fact that the differential equation is none other than the Legendre equation for $P_{1/2}(x)$. Once the coefficients for $a_k$, $b_k$ and $c_k$ are computed, the asymptotic value for the pressure within the liquid domain can be obtained from Eq. (\ref{eq:pressure_selfsimilar}) for $\partial \psi/\partial t = 0$ by collecting terms order by order:
\begin{align}
&p(\textsf{r} \to \infty, \theta \geq \theta_0)= \nonumber\\
&\frac{1}{\textsf{r}}\left\{a_1-\frac{a_0^2}{8}\left[4 \sin^2\theta P^{'~2}_{{1}/{2}}(-\cos\theta)+P^2_{1/2}(-\cos\theta)\right]\right\}\nonumber\\
&+\frac{1}{\textsf{r}^{5/2}}\left[\frac{a_0 a_1}{2}P_{1/2}(-\cos \theta)+2a_2 P_{3/2}(-\cos\theta)\right] \nonumber\\
&+\mathcal{O}(\textsf{r}^{-4}) \label{eq:asymptoticPressure}.
\end{align}
Although the pressure contains the sink flow contribution  $\mathcal{O}(1/r)$, there are now additional contributions dependent on the polar coordinate $\theta$, which for certain choices of parameter values can generate a ballistic unidirectional flow, as depicted in Fig. \ref{fg:comparesolns}(c).	

\begin{figure}
\includegraphics[scale = 1]{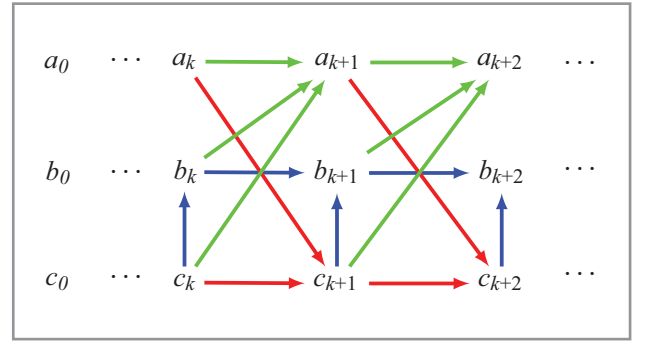}
\caption{Flow chart depicting the interweave process for computing the coefficients $a_k$, $b_k$, $c_k$. Arrows represent the equations required for computing the coefficient shown to the next higher order as required by simultaneous solutio of the equipotential condition in Eq. (\ref{eq:SSequipot}) (blue arrows), the kinematic boundary condition in Eq. (\ref{eq:SSkin}) (red arrows) and the Bernoulli equation given by Eq. (\ref{eqn:SphericBern}) (green arrows).
\label{fg:abcFlowChart}}
\end{figure}

\subsection{Estimation of near field conic behavior}
\label{sec:nearfield}

\begin{figure}[!]
\includegraphics[scale=1]{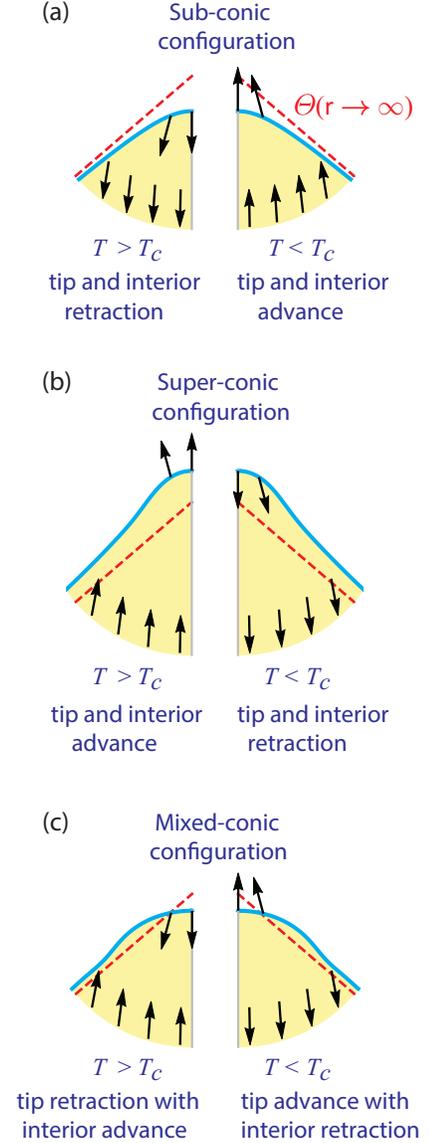}
\caption{Allowable configurations showing relative motion between fluid in the apical tip and nearby bulk fluid. Dashed line (red) represents the Taylor cone envelope. Curved lines (light blue) denote the liquid interface. (a) Sub-conic configurations in which both the tip and bulk fluid together advance or recede from the cone envelope from below. (b) Super-conic configurations in which both the tip and bulk  fluid together advance or recede from the cone envelope from above. (c) Mixed-conic configurations in which the tip and bulk fluid move in opposite directions with some portions of the interface lying below and some above the Taylor cone envelope. This configuration describes cases in which the liquid interface lies above the Taylor cone envelope, as here depicted, or below the envelope (not shown). }
\label{fg:supsubmix}
\end{figure}

Insight into the various flow configurations which can occur in the vicinity of the liquid tip can be obtained with reference to the point of intersection between the liquid apex as measured on the central axis of symmetry and the sloped line defining the tangent to the asymptotic liquid surface specified by $\theta_0$. To begin, we note that due to symmetry, the kinematic condition in Eq. (\ref{eqn:ZubarevKinematic}) when evaluated at the liquid apex $[r=0,z =h(0)]$ reduces simply to
\begin{equation}
\psi_z \Big|_{\textrm{apex}} = -\frac{2}{3}h_{\textrm{apex}}.
\end{equation}
The apex fluid velocity is therefore controlled by the sign of $h(0)$, the value of the projection of the liquid height function on the central axis. That is, when for $T<T_C$ the fluid approaches the intersection point from below, then $z_\textrm{apex} = h(0)<0$ and the velocity is positive and the fluid moves upward. Similarly when for $T>T_C$ and the fluid approaches the intersection point from above, then $z_\textrm{apex} = h(0)>0$ and the velocity is negative.  Mixed behavior should also be possible in which the fluid in the apical region is moving upward but the interior fluid is moving downward or vice versa, as occurs in capillary pinch-off phenomena  \cite{Sierou04,Hoepffner13} due to inertial effects. Since the inviscid form of the Bernoulli equation is invariant under time-reversal symmetry, we anticipate several categories of flow as illustrated in Fig. \ref{fg:supsubmix}, which we coin sub-conic, super-conic and mixed-conic.

In his original analysis, Zubarev \cite{Zubarev01} developed an elegant argument for the distance between the liquid apex and the apex of the conic envelope based on the fact that the velocity potential is a harmonic function and therefore the net normal flux  corresponding to $\nabla \psi$ when integrated across any closed surface must vanish. What simplified the analysis in that case was that the correction to the leading order term in the expansion for $h - (\cot\theta_0) \textsf{r} \sim \mathcal{O}(r^{-5})$. A similar approach applied to our case where $h(\textsf{r},\theta) - c_0\textsf{r} = c_1\textsf{r}^{-1/2}+...$ introduces some  difficulties. By invoking some minor additional assumptions, however, we show it is possible to obtain an analytic result.

\begin{figure}[!]
\includegraphics[scale=1]{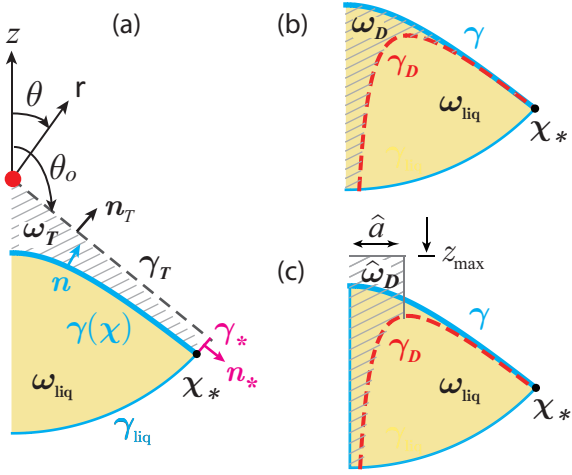}
\caption{Sketches showing the geometry for estimating the liquid apex height along the central axis of symmetry. Volumes indicated by $\omega_T$, $\omega_D$ and $\hat{\omega}_D$ can either be positive or negative depending on whether the liquid is approaching the conic envelope intersection point (red) from below or above. (a) The domain $\omega_\mathrm{T}$ (hatched lines) with outwardly pointing unit normal vector $\vec{n}_T$ represents the interstitial volume between the free surface liquid boundary $\gamma$ (curved blue line) encapsulating the volume $\omega_\textrm{liq}$ and the Taylor-conic envelope $\gamma_T$ (dashed black line) with polar angle $\theta_0$. The coordinate $\vec{\x}_*=(\textsf{r}_*,\theta_0)$ represents the liquid domain terminus point. The boundary element $\gamma_*$ (magenta line) represents a small circular extension to the radial boundary $\gamma_liq$ - the unit normal $\vec{n}_*$ is therefore oriented along the radial axis. (b) The domain $\omega_D$ (hatched lines) represents the interstitial volume between the liquid surface boundary $\gamma$ and the surface boundary $\gamma_D$ (dashed red line), whose outer branch represents the leading order asymptote $z=c_0 \rc +c_1 \rc^{-1/2}$. (c) The domain $\hat{\omega}_D$ (hatched lines) represents the volume encapsulated between the surface boundary $\gamma_D$ below, the horizontal axis $z_max$ above, the central axis to the left and the distance $\hat{a}$ to the right, where $\hat{a}$ denotes the lateral distance corresponding to the maximum of the boundary $\gamma_D$. The quantity $z_\mathrm{max}$ denotes the maximum elevation point of  $\gamma$ along the central axis.
\label{fg:integralConstraint}}
\end{figure}

We refer to the representation in Fig. \ref{fg:integralConstraint}(a)  in developing an estimate for the interstitial volume $\omega_T$, which is bounded by the axis of symmetry, the free surface boundary $\gamma(\vec{\x})$, the small circular arc of radius $\textsf{r}_*$ and unit normal $\vec{n}_*$, and the Taylor conic envelope $\gamma_T$ defined by the polar angle $\theta_0 = \pi - \theta_T$. The coordinate $\vec{\x}_*$ designates a truncation point on $\gamma(\vec{\x})$ such that $|\textsf{r}_*| \gg 1$. The vertical separation distance between $\gamma_T$ and $\gamma$ can be positive or negative depending on whether $\gamma(\vec{\x})$ lies below (sub-conic configuration) or above (super-conic configuration) the boundary $\gamma_T$, as depicted in Fig. \ref{fg:supsubmix}. Consequently, the volume $\omega_T$ can be positive or negative depending on whether the liquid is approaching the conic envelope intersection point (red) from below or above, respectively.
The interstitial volume $\omega_T$ can be computed from the relation
\begin{align}
&\omega_T=\int_{\omega_T}d\omega=\frac{1}{3}\int_{\omega_T}\nabla\cdot \vec{\x}~d\omega \nonumber\\
&=\frac{1}{3}\int_{S_T}(\vec{\x} \cdot \vec{n}_T)~dS+\frac{1}{3}\int_{S_*} \vec{n}_*\cdot \vec{\x}~ dS-\frac{1}{3}\int_{S_\gamma}\vec{n}\cdot \vec{\x}~dS~,
\label{eq:volBetween}
\end{align}
where $dS$ refers to the surface boundary element and $S_T$, $S_*$ and $S_\gamma$ refer to the areas of the bounding surfaces - no fluid can penetrate the central axis due to axisymmetry and so that area does not contribute to the sum. The negative sign in the last term reflects the fact that the unit normal vector pointing outward from the volume $\omega_T$ along the boundary $\gamma$ equals $-\vec{n}$. Since on the conic surface, $\vec{\x} \cdot \vec{n}_T=0$, the corresponding integral in Eq. (\ref{eq:volBetween}) vanishes. The integral over $S_*$ reduces to $(2\pi/3)\,\textsf{r}^2_*\, [\textsf{r}_*\cos\theta_0 - h(\textsf{r}_*,\theta_*)]$, where $h$ is the projection of the terminus point $\vec{\x}_*$ onto the central axis. For the configuration depicted in Fig. \ref{fg:integralConstraint}(a), for example, $h<0$. Evaluation of the integral over $S_\gamma$ requires an analytic expression for $\gamma$. To proceed, we invoke the kinematic boundary condition in Eq. (\ref{eq:unsteadyKinematicSS}) with $\partial f/\partial t=0$. Application of the divergence theorem along with the observation that  $\nabla^2 \psi = 0$ within $\omega_\textsf{liq}$ yields
\begin{align}
&-\frac{1}{3}\int_{S_\gamma}\vec{n}\cdot\vec{\x}~dS= \frac{1}{2}\int_{S_\gamma} \vec{n}\cdot\nabla\psi~ dS \nonumber\\
&= \frac{1}{2}\left(\int_{S_\gamma \cup S_\textrm{liq}} - \int_{S_\textrm{liq}}\right) \vec{n}\cdot\nabla\psi ~dS \nonumber \\
&= \frac{1}{2}\int_{\omega_\textrm{liq}} \nabla \cdot \nabla \psi~ d\omega - \frac{1}{2}\int_{S_\textrm{liq}} \vec{n}\cdot\nabla\psi ~dS \nonumber\\
&=- \pi \int_{S_\textrm{liq}}\vec{n}\cdot\nabla\psi ~\textsf{r}^2_* \sin\theta ~d\theta~,
\label{eq:integralLiqSphericalCap}
\end{align}
where the quantity $S_\gamma \cup S_\textrm{liq}$ denotes the union of the areas indicated which encompass the volume $\omega_\textrm{liq}$. For sufficiently large $\textsf{r}$, the velocity potential $\psi$ on $S_\textrm{liq}$ approaches its  asymptotic value $\phi_\infty$ given by the expansion in Eq. (\ref{eq:psiexpansion}), which when substituted into Eq. (\ref{eq:integralLiqSphericalCap}) yields
\begin{align}
&\mathbin{\phantom{=}} \int_{S_\textrm{liq}} \vec{n}\cdot\nabla\psi ~ \textsf{r}^2_* \sin\theta ~ d\theta \nonumber\\
&=\int^\pi_{\theta_*}\pd{\psi_\infty}{\textsf{r}}\Big\rvert_{(\textsf{r}_*,\theta)}
\textsf{r}^2_* \sin\theta ~d\theta \nonumber\\
&=\sum_{k=0}^\infty a_k\int^\pi_{\theta_*}\pd{\psi_k}{\textsf{r}}\Big\rvert_{(\textsf{r}_*,\theta)}
\textsf{r}^2_* \sin\theta ~ d\theta \nonumber\\
&=\sum_{k=0}^\infty a_k\int^\pi_{\theta_*}
\pd{r^{\frac{1}{2}-\frac{3}{2}k}}{\textsf{r}}
\Big\rvert_{r_*}
P_{\frac{3}{2}k-\frac{3}{2}}(-\cos\theta)
~\textsf{r}^2_* \sin\theta ~ d\theta \nonumber\\
&=\sum_{k=0}^\infty a_k
\frac{1-3k}{2}\textsf{r}^{\frac{3}{2}-\frac{3}{2}k}_*
\int^\pi_{\theta_*}P_{\frac{3}{2}k-\frac{3}{2}}(-\cos\theta) \sin\theta ~d\theta \nonumber\\
&=\sum_{k=0}^\infty a_k
\frac{1-3k}{2} \textsf{r}^{\frac{3}{2}-\frac{3}{2}k}_*
\int^{1}_{-\cos\theta_*}P_{\frac{3}{2}k-\frac{3}{2}}(x)~dx \nonumber\\
&=-a_1(1+\cos\theta_*)+ \frac{2a_0}{3}\textsf{r}^{3/2_*}
\sin^2\theta_*
P'_{\frac{1
}{2}}(-\cos\theta_*)\nonumber\\
&\mathbin{\phantom{=}}+\sum_{k=2}^\infty a_k
\textsf{r}^{\frac{3}{2}-\frac{3k}{2}}_*
\frac{2\sin^2\theta_*}{3(1-k)}
P'_{\frac{3k}{2}-\frac{3}{2}}(-\cos\theta_*)~.
\label{eq:integratedkinematic}
\end{align}
The last step is obtained by noting the relation
\begin{equation}
\int_x^1P_\nu(x')\,\mathrm{d}x'=\frac{1-x^2}{\nu(\nu+1)}P_\nu'(x) \quad \quad \textrm{for}~ \nu\neq 0~,
\end{equation}
which is obtained by integration of the Legendre differential equation. Substitution of Eq. (\ref{eq:integralLiqSphericalCap}) and Eq. (\ref{eq:integratedkinematic}) into Eq. (\ref{eq:volBetween}) results in the series solution
\begin{align}
&\omega_T=a_1\pi (1+\cos\theta_*)-a_0\frac{2\pi \sin^2\theta_*}{3}
\textsf{r}^{3/2}_* P'_{1/2}(-\cos\theta_*)\nonumber \\
&+...+\frac{2\pi}{3}\textsf{r}^2_*\left[\textsf{r}_*\cos\theta_0 - h(\textsf{r}_*,\theta_*)\right]~.
\label{eq:omegaTsum}
\end{align}
We now expand each term in powers of $\textsf{r}$, keeping in mind that $c_2 = 0$ in Eq. (\ref{eq:hexpansion}). After some straightforward algebra, we find
\begin{align}
&\omega_T=a_1\pi (1+\cos\theta_0) +\frac{a_0^2\pi}{3}P_{1/2}'(-\cos\theta_0) \nonumber \\
&~~~~~~~~~\times\left[\frac{\cos\theta_0P_{1/2}'(-\cos\theta_0)}{(1+\cos^2\theta_0)^{3/2}}+\frac{ 2 P_{1/2}''(-\cos\theta_0)}{(1+\cos^2\theta_0)^2}\right] \nonumber\\
&-\int_0^{\rc_*}c_1\rc^{-1/2}\,2\pi r dr +O(r^{-3/2}_*)~.
\label{eq:signVolDecayResidue}
\end{align}
Since all terms are bounded as $\textsf{r}_* \to \infty$, the final result yields
\begin{align}
\label{eq:TCintegralConstraint_v0}
&\lim_{{r}_*\to \infty}\left\{\int_{\omega_T}~d\omega+\int_0^{\textsf{r}_*}c_1~r^{-1/2}_*~2\pi \textsf{r }d\textsf{r}\right\}\nonumber\\
&=a_1\pi (1+\cos\theta_0)
+\frac{a_0^2\pi}{3} P_{1/2}'(-\cos\theta_0) \nonumber\\
&~~~~~~~~\times\left[\frac{\cos\theta_0P_{1/2}'(-\cos\theta_0)}{(1+\cos^2\theta_0)^{3/2}} +\frac{ 2 P_{1/2}''(-\cos\theta_0)}{(1+c_0^2)^2}\right].
\end{align}
We see from Eq. (\ref{eq:a0a1a2}) and Eq. (\ref{eq:c0c1c2}) that the parameter choice $a_0 = c_0 = 0$ when substituted into Eq. (\ref{eq:TCintegralConstraint_v0}) reduces to the solution first obtained by Zubarev \cite{Zubarev01} - namely, $\psi \sim r^{-1}$. By contrast, the derivation outlined above yields $\psi \sim a_0r^{1/2} + a_1 r^{-1}$, in agreement with his subsequent finding \cite{Suvorov04}.

A more intuitive geometric interpretation of the quantity in Eq. (\ref{eq:TCintegralConstraint_v0})
can be had by introducing a new asymptotic boundary $\gamma_\mathrm{D}$ parameterized by coordinates $(\textsf{r}, z=\textsf{r}\cos\theta)$, as depicted in Fig. \ref{fg:integralConstraint}(b), where
\begin{equation}
\gamma_D = \{(\textsf{r}, \cot\theta_0 \textsf{r} +c_1 \textsf{r}^{-1/2})\mid 0 \le \textsf{r} \le \textsf{r}_*\},
\label{eq:gammaD}
\end{equation}
and a corresponding volume $\omega_D$ defined by integration over the difference in vertical height between the projection of $\gamma_D$ onto the central axis and that of $\gamma$ as defined by $h$. Accordingly, we find
\begin{align}
\label{eq:TCintegralConstraint_v1}
&\lim_{\textsf{r}_*\to \infty}\omega_D
=\lim_{\textsf{r}_*\to\infty}\int_0^{\textsf{r}_*}\left[\cot\theta_0\textsf{r} + c_1\textsf{r}^{-1/2}-h(\textsf{r},\theta)\right]2\pi \textsf{r} d\textsf{r} \nonumber\\
&=a_1\pi (1+\cos\theta_0) +\frac{a_0^2\pi}{3} P_{1/2}'(-\cos\theta_0) \nonumber\\
&~~~~~~~~~~~~~~~~\times \left[\frac{\cot\theta_0 P_{1/2}'(-\cos\theta_0)}{(1+\cos^2\theta_0)^{3/2}} +
\frac{ 2 P_{1/2}''(-\cos\theta_0)}{(1+\cos^2\theta_0)^2}\right],
\end{align}
which when evaluated numerically yields $\lim_{{r}_*\to \infty}\omega_\mathrm{D} \approx 1.093\, a_1-0.118\, a_0^2$.
Because of the recursion relations specified by Eqs. (\ref{eq:a0_a1_b0}) and (\ref{eq:c0c1c2}), Eq. (\ref{eq:TCintegralConstraint_v1}) therefore poses a constraint between $a_1$ and $c_1$, or equivalently between  $a_0$ and $b_0$.  Arbitrary combinations of these pairs are not therefore necessarily admissible.

From Eq. (\ref{eq:TCintegralConstraint_v1}), the maximum distance between the liquid apex and the conic envelope intersection point can now be estimated. For simplicity, we restrict our attention to the case $c_1 < 0$ and a boundary $\gamma$ that is everywhere bounded below by $\gamma_\mathrm{D}$, according to which
\begin{equation}
\label{eq:assumption}
\cos\theta_0 \textsf{r} +c_1 \textsf{r}^{-1/2} \le  h(\textsf{r},\theta) \le \cot\theta_0 \textsf{r}
~~~\textrm{for}~r\ge 0 ~\textrm{and}~ c_1<0.
\end{equation}
Referring to Fig. \ref{fg:integralConstraint}(c), the length $\hat{a}$ defines the projected distance onto the central axis of the turning point where the boundary $\gamma_D$ attains its maximum amplitude. The maximum distance $z_\mathrm{max}$ between the liquid apex and the point of intersection with the conic envelope can then be obtained from the requirement that the volume $\hat{\omega}_D$, which is bounded between $z=z_\mathrm{max}$ and $\gamma_\mathrm{D}$ up through the distance $\hat{a}$ shown is identical to $\omega_\mathrm{D}$ as $\lim_{\textsf{r}_*\to \infty}$:
\begin{equation} \label{eq:TCintegralConstraint_v2}
\hat{\omega}_\mathrm{D}=\int_0^{\hat{a}} \left(
\cos\theta_0 \textsf{r} + c_1 \textsf{r}^{-1/2} - z_\mathrm{max} \right) \,2\pi r~dr
=\lim_{\textsf{r}_*\to \infty}\omega_\mathrm{D}~.
\end{equation}
A straightforward exercise then yields the following expressions for the quantities shown:
\begin{align}
\hat{a}&=\left(\frac{c_1}{2\cos\theta_0}\right)^{2/3} \\
z_\mathrm{max}&=\left(\frac{2 \cos\theta_0}{c_1}\right)^{1/3}\left(
\frac{5}{3}  c_1-\frac{2  \cos\theta_0 }{\pi  c_1}\lim_{\textsf{r}_*\to \infty}\omega_D
\right).
\end{align}
It follows immediately from evaluation of Eq.  (\ref{eq:assumption}) to the interval $0\le \textsf{r} \le \hat{a}$ that the maximum elevation $h$ must occur above the maximum elevation of $\gamma_D$ at $\hat{a}$, and the minimum elevation of $h(r,\theta)$ must occur below $z_\mathrm{max}$:
\begin{align}
\label{eq:TCintegralConstraint_v3}
\frac{3c_1}{2^{2/3}} \left(\frac{\cos\theta_0}{c_1}\right)^{1/3}
&\le \max_{\textsf{r}\in[0,\hat{a}]}h(\textsf{r},\theta) \quad \textrm{and} \\
\min_{\textsf{r}\in[0,\hat{a}]}h(\textsf{r},\theta) &\le z_\mathrm{max}.
\end{align}
These bounds differ from the original result by Zubarev \cite{Zubarev01} since here, the leading asymptotic term in $\psi$ scales as $r^{1/2}$ and not $r{-1}$, which leads to a liquid height bounded below by the projected coordinate  $z=\cos\theta_0 r+c_1 /\sqrt{r}$.

\section{Boundary Integral Patching Technique For Complete Numerical Solutions}
\label{sc:NumSolPatBIE}

The asymptotic self-similar solutions so far obtained for the velocity potential, electric potential and interface shape describe  only the functional behavior at distances far from the liquid apex. The near field behavior is not amenable to analytic solution and therefore requires a fully numerical approach. In what follows, we describe implementation of a boundary integral patching technique which utilizes as boundary data those series expansions in Section \ref{sec:dynselfsim} representing the asymptotic self-similar solutions. For additional details on the method of implementation, the interested reader may wish to consult Ref, [\onlinecite{Leppinen03}] for a similar implementation in investigating the problem of capillary pinch-off.

\begin{figure}[!]
\includegraphics[scale=1]{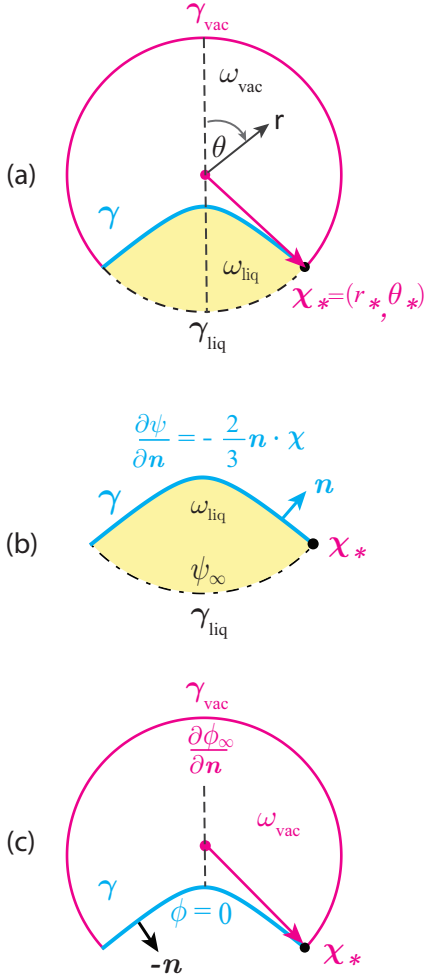}
\caption{Sketches illustrating the two partitions comprising the truncated circular domain of radius $\textsf{r}_*\gg 1$ relevant to numerical computation. Because of spherical symmetry about the central vertical axis (dashed grey line), the actual computational domain consisted only of the areas and boundaries to the right of the axis. (a) Liquid domain volume $\omega_\textrm{liq}$ bounded above by the boundary $\gamma$ (teal solid line) and below by the boundary $\gamma_\textrm{liq}$ (black dashed line). Vacuum domain volume $\omega_\textrm{vac}$ bounded above by the boundary $\gamma_\textrm{vac}$ (magenta solid line) and below by the boundary $\gamma$. Coordinates of the domain truncation point specified by  $\vec{\x}_* = (\textsf{r}_*, \theta_*)$. (b) Kinematic boundary condition for $\psi$ from Eq. (\ref{eq:trunckinematic}) applied to boundary points on $\gamma$. Boundary values for $\psi_\infty$ from Eq. (\ref{eq:truncpsi}) applied to boundary points on $\gamma_\textrm{liq}$. (c) Equipotential condition $\phi=0$ from Eq. (\ref{eq:truncequip}) applied to boundary points on $\gamma$. Boundary values for $\partial \psi_\infty/\partial \vec{n}$ were extracted from Eq. (\ref{eq:truncphi}) and applied to boundary points on $\gamma_\textrm{vac}$.}
\label{fg:bieTrunc}
\end{figure}

Shown in Fig. \ref{fg:bieTrunc} are illustrations of the liquid and vacuum domain volumes inscribed by a circular boundary of radius $\textsf{r}_*\gg 1$ used for numerical solutions, along with the  boundary conditions applied.
Since the solutions for the velocity potential $\psi$ and electric potential $\phi$ are harmonic functions, it is known from Green's theorem that they can therefore be represented by the boundary integrals
\begin{align}
\label{eq:bie}
&\beta(\vec{\x}')\psi(\vec{\x}')= \nonumber \\
&\int_{S_\gamma \cup S_\textrm{liq}}
\left\{\mathdutchcal{g}(\vec{\x};\vec{\x}')\pd{\psi(\vec{\x})}{\vec{n}}-
\psi(\vec{\x})\pd{\mathdutchcal{g}(\vec{\x};\vec{\x}')}{\vec{n}}
\right\} dS\\
&\beta(\vec{\x}')\phi(\vec{\x}')= \nonumber \\
&\int_{S_\gamma \cup S_\textrm{vac}}
\left\{\mathdutchcal{g}(\vec{\x};\vec{\x}')\pd{\phi(\vec{\x})}{\vec{n}}-
\phi(\vec{\x})\pd{\mathdutchcal{g}(\vec{\x};\vec{\x}')}{\vec{n}}
\right\} dS~,
\label{eq:bie}
\end{align}
where $\beta$ represents the interior angle between two adjacent segments connected at $\vec{\x}'$ along the discretized boundary. The axisymmetric Green's function expressed in spherical coordinates is given by
\begin{equation}
\mathdutchcal{g}(\vec{\textsf{r}};\textsf{r}') = \int^{2\pi}_0 \mathdutchcal{G}(\vec{\textsf{r}};\vec{\textsf{r}}')~d\varphi~,
\end{equation}
where $\varphi$ is the spherical azimuthal angle, $\nabla^2 \mathdutchcal{G}(\vec{\textsf{r}};\vec{\textsf{r}}')=4\pi \delta (|\vec{\textsf{r}} - \vec{\textsf{r}}'|)$, $\delta$ is the Dirac delta function and $\mathdutchcal{G}(\vec{\textsf{r}};\vec{\textsf{r}}')= |\vec{r} - \vec{\textsf{r}}'|^{-1}$. The axisymmetric function $\mathdutchcal{g}(\vec{\textsf{r}};\vec{\textsf{r}}')$, which represents the strength of the velocity potential at $\vec{\textsf{r}}$ from a ring source located at $\vec{\textsf{r}}'$, cannot be expressed in terms of elementary functions. Additional details pertaining to the calculation and numerical evaluation of simpler two-dimensional boundary integral problems can be found in Ref. [\onlinecite{Lennon79}]. We note that while in many boundary value problems the contributions from far field boundaries typically decay rapidly and can be neglected, that is not the case here. For example, when evaluated along the boundary segment $\gamma_\textrm{liq}$, where $\psi$ adopts the value $\psi_\infty$ given by Eq. (\ref{eq:psiexpansion}), the integral contribution
\begin{equation}
\int_{S_\textrm{liq}}
\psi(\vec{\x})\pd{\mathdutchcal{g}(\vec{\x}';\vec{\x})}{\vec{n}}~
dS~ \sim \textsf{r}^{1/2}_* ~,
\end{equation}
cannot be neglected since it scales as $\textsf{r}^{1/2}_*$. The same scaling occurs for the term in the expression for the electric potential involving the terms $\mathdutchcal{g}$ and $\partial \phi/\partial \vec{n}$. Contributions arising from the boundaries $\gamma_\textrm{liq}$ and $\gamma_\textrm{vac}$ need therefore be included in the computation. For completeness, the following closure conditions were therefore applied along the specified boundary segments:
\begin{align}
\frac{\partial\psi}{\partial\vec{n}}&=-\frac{2}{3}\vec{n}\cdot\vec{\x} &\textrm{on}~\gamma \label{eq:trunckinematic}\\
\psi_\infty &= \sum^4_{k=0}a_k~ \textsf{r}^{\frac{1}{2}-\frac{3}{2}k}~P_{\frac{3}{2}k-\frac{3}{2}}(-\cos\theta) &\textrm{on}~\gamma_\textrm{liq} \label{eq:truncpsi}\\
\phi&=0 &\textrm{on}~\gamma \label{eq:truncequip}\\
\phi_\infty &= \sum^4_{k=0}b_k ~ \textsf{r}^{\frac{1}{2}-\frac{3}{2}k}~P_{\frac{3}{2}k-\frac{3}{2}}(\cos\theta) &\textrm{on}~\gamma_\textrm{vac}\label{eq:truncphi}\\
\end{align}

\begin{figure}[!]
\includegraphics[scale=1]{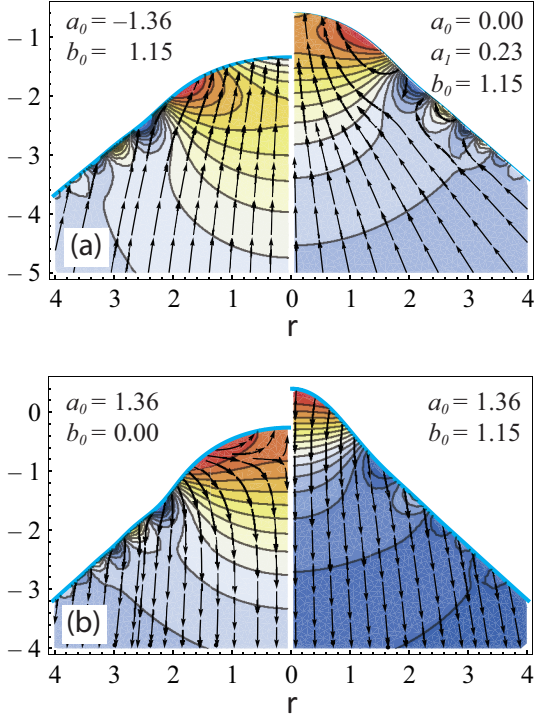}
\caption{Exact numerical solutions obtained by the boundary integral patching technique showing the free surface liquid shape (solid teal lines), liquid isobaric contours (solid dark grey lines) and velocity field (black arrows) for the parameter values indicated. (a) Left image: Sub-conic configuration for $b_0=1.15$ and $a_0 = -1.36$. Right image: Super-conic configuration for $b_0=1.15$ and $a_0=0$ or  $a_1=0.23$. (b) Left image: Mixed-conic configuration for $a_0 = 1.36$ and $b_0=0.00$. Right image: Super-conic configuration for $a_0 = 1.36$ and $b_0=1.15$.}
\label{fg:numresults}
\end{figure}

The boundary points defining $\gamma$, $\gamma_\textrm{liq}$ and $\gamma_\textrm{vac}$ were interpolated by quintic splines. Quadratic Lagrange basis functions along the spline arc-length were used to approximate $\psi$, $\partial\psi/\partial\vec{n}$, $\phi$ and $\partial \phi/\partial \vec{n}$ using the Lagrange nodal values along the boundary. For integrals involving  $\mathdutchcal{g}(\vec{\chi}:\vec{\chi}')$ in which the singular point $\vec{\textsf{r}}=0$ was not located at either end of a discretized boundary element, the Green's function and its normal derivative behave regularly and so integration proceeded by Gauss-Legendre quadratures. For integrals involving  $\mathdutchcal{g}(\vec{\chi}:\vec{\chi}')$ in which the singular point $\vec{\textsf{r}}=0$ occurred at either end of a discretized boundary element, the normal derivative of the Green's function has a logarithmic singularity and so integration relied instead on logarithmic-weighted quadratures.

The numerical scheme then proceeded as follows. Given the parameter pair ($a_0$, $b_0$) for $\psi$ and $\phi$ specified by Eq. (\ref{eq:psiexpansion}) and Eq. (\ref{eq:phiexpansion}), a corresponding trial interface function and its first two derivatives was obtained from Eq. (\ref{eq:hexpansion}) and evaluated at the truncation point $\vec{\x}_*$. The matrix equations resulting from evaluation of the discretized integral equations for $\psi$ and $\phi$ from Green's identify and subject to the boundary conditions in Eq. (\ref{eq:trunckinematic}) - (\ref{eq:truncphi}) were then solved by QR decomposition to obtain the nodal values for $\psi$ and $\partial \phi /\partial \vec{n}$ on $\gamma$. A Newton-Raphson method was then applied iteratively to adjust parametrization of the interface $\gamma$ until $\psi$ and $\phi$ satisfied the time- independent form of Bernoulli's equation given by Eq. (\ref{eq:unsteadyBernoulliSS} (with $\partial \psi/\partial t$). The  Jacobian matrix at each Newton step was numerically
approximated by perturbations to $\gamma$ along the boundary normals.

Plotted side-by-side in Fig. \ref{fg:numresults}(a) are the exact numerical solutions from the patched boundary integral technique for the same strength of the far-field electric potential $b_0$ but different values of the leading velocity potential coefficient $a_0$. The left image in Fig. \ref{fg:numresults}(a) depicts a case in which the liquid surface lies below the Taylor cone envelope. The velocity field here is not radially oriented but closely aligned with the vertical axis due to the leading order ($k=0$) angular dependence of $\psi$ in Eq. (\ref{eq:truncpsi}).

According to Fig. \ref{fg:supsubmix}(a), this represents a sub-conic configuration with tip and interior fluid advance for $T<T_C$. Under time reversal symmetry such that $T>T_C$, the velocity field reverses and so this flow profile describes equivalently a sub-conic configuration with tip and interior fluid retraction. The solution in the right  image of Fig. \ref{fg:numresults}(a) depicts a case in which the liquid surface lies above the Taylor cone envelope. Except within a short distance of the interface, the velocity field is radially oriented. According to Fig. \ref{fg:supsubmix}(b), this represents a super-conic configuration with tip and interior fluid advance for $T>T_C$, or, from time reversal symmetry, a super-conic configuration with tip and interior fluid retraction for $T<T_C$. Both examples in Fig. \ref{fg:numresults}(a) also exhibit multiple stagnation points along the interface, as indicated by the small and nested semi-circular isobaric contours.

Plotted side-by-side in Fig. \ref{fg:numresults}(b) are results  obtained using the same leading velocity potential coefficient $a_0$  but different values of the electric field potential coefficient $b_0$. The flow profile in the left image corresponds to a mixed-conic configuration in which the region near the liquid apex advances toward the Taylor envelope from below while the interior fluid undergoes retraction for $T<T_C$, or likewise a mixed-conic configuration with tip retraction and interior advance for $T>T_C$. The choice $b_0 = 0$ corresponds to the absence of an applied electric field and so the configuration shown reflects liquid motion solely acted upon by capillary and inertial forces. The fact that the  velocity field for this choice of parameters exhibits a spherical cap region that is advancing while the interior flow is retracting is indicative of flow precursors that ultimately lead to capillary pinchoff. These results are very similar to those previously obtained reported for post-pinchoff recoil behavior after droplet elongation \cite{Sierou04}. The right image in Fig. \ref{fg:numresults}(b) corresponds to a super-conic configuration with tip and interior fluid retraction for $T<T_C$, or equivalently, tip and interior advance for $T>T_C$. Both solutions in Fig. \ref{fg:numresults}(b) also exhibit multiple stagnation points along the interface.

For decades, researchers have been interested in quantifying the influence of charge transport on the shape and emission of progeny drops during Coulombic fission. For a perfectly conducting mass like a liquid metal, the electrical conductivity is assumed infinite and any excess charge on the surface redistributes instantaneously in order to maintain equipotential conditions. The charge density and electric field then depend purely on geometry. In a seminal paper in 2011, Burton and Taborek \cite{Burton11} simulated the deformation accompanying Coulombic fission of an isolated inviscid droplet of perfectly conducting liquid of density $\rho_liq$ of sufficiently high surface charge density embedded within an exterior fluid of lower density $\rho_\textrm{ext} = 0.001 \rho_\textrm{liq}$.  Their numerical simulations revealed formation of self-sharpening tips at the poles in which the apical values of the surface charge density, interface curvature, and velocity underwent divergence in finite time (with no progeny droplets observed). More importantly, they uncovered robust power law growth in the apical fields spanning an incredible 12 decades in time according to which the apex curvature scaled as $0.604\, \tau^{-2/3}$ and the surface charge density (or equivalently the electric field strength) scaled as $0.925~ \tau^{-1/3}$.
\begin{figure}[!]
\includegraphics[scale=1]{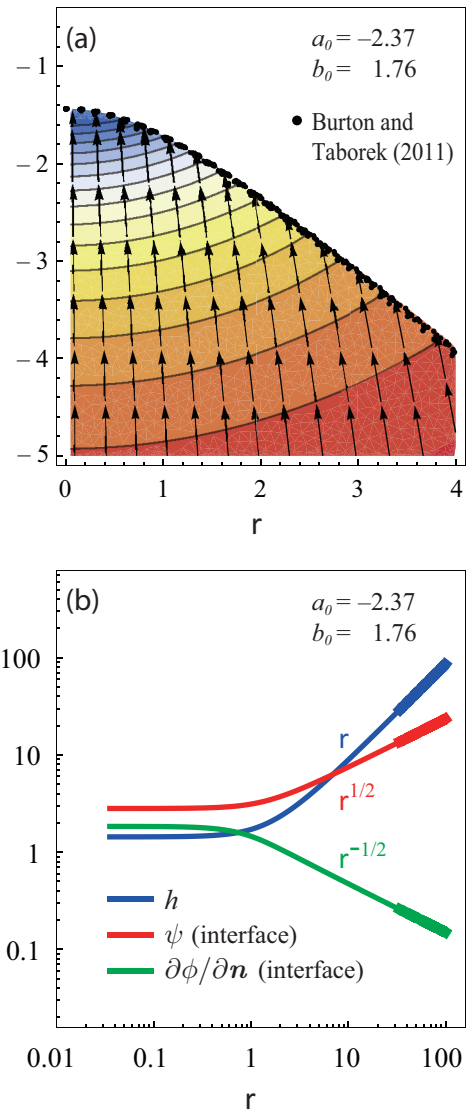}
\caption{(a) Exact numerical solutions from the boundary integral patching technique showing the shape of the liquid interface, isobaric contours within the liquid (solid dark grey lines) and velocity field within the liquid (black arrows) for $a_0 = -2.37$ and $b_0=1.76$. The superposed dots (black) along the liquid interface are the data points extracted from the simulations of Burton and Taborek \cite{Burton11}. (b) Numerical solutions from the boundary integral patching technique for the projected height $h(\textsf{r})$ (blue line), surface velocity potential $\psi$ (red line) and surface  electric field strength $\partial \phi /\partial \vec{n}$ (green line) as a function of $\textsf{r}$. The thicker line segments at large $\textsf{r}$ represent the leading order solutions ($k=0$) evaluated along the liquid interface from Eqs. (\ref{eq:psiexpansion}) - (\ref{eq:hexpansion}) with coefficient values specified by Eqs. (\ref{eq:a0a1a2}) - (\ref{eq:c0c1c2}), namely $\psi_\infty = 0.989703~a_0~ \textsf{r}^{1/2}$, $\partial \phi^\infty/\partial \textsf{r} =0.848582~b_0~\textsf{r}^{-1/2}$ and $h_\infty=c_0\,\textsf{r}$ for $a_0=-2.37$, $b_0 = 1.76$ and $c_0= 0.860437$.}
\label{fg:compareBT}
\end{figure}

To make contact with their simulations, we magnified the image in Fig. 1(b) of Ref. [\onlinecite{Burton11}] showing numerous snapshots in time of a liquid tip evolving into a dynamic cone. We extracted the boundary data points for 20 such snapshots and transformed these  points to the self-similar frame. Though no length scale accompanied their plot, we determined the correct overall isotropic scale by conducting a least squares fit between their (transformed) data sets and ours for coefficient values $a_0 =2.37$ and $b_0 =1.757$. This comparison led to a value of the liquid apex curvature of $-0.608$ and a value for the liquid apex electric field strength of $0.922$. Shown in Fig. \ref{fg:compareBT}(a) are the self-similar data points extracted from Burton and Taborek \cite{Burton11} superposed on the interface shape computed from our boundary integral technique. The agreement is excellent despite that in their simulations, the exterior fluid was a liquid of low density and not vacuum as in our model. Shown in Fig.\ref{fg:compareBT} (b) are our numerical results for the radial dependence of the projected height, surface velocity potential and surface electric field strength with increasing $\textsf{r}$.

\section{Conclusion}
In this work, we have quantified the dynamic behavior accompanying the accelerated evolution of an axisymmetric protrusion in an electrically stressed, perfectly conducting liquid subject to irrotational flow. The analysis relies on the inviscid Bernoulli equation applied to the moving interface, which includes not only capillary and Maxwell forces but the critical influence of inertial forces. The inviscid approximation is appropriate to studies of liquid metals since the viscous boundary layer extends only tens of nanometers from the moving interface. Given the complexity of the coupled equations and boundary conditions required to fully describe the electrohydrodynamic behavior, the only analytic solutions possible are those asymptotic expansions in the far field at distances $\textsf{r}$ large from the liquid apex. These series expansions for the velocity potential, electric field potential and interface are represented by a two-parameter family of self-similar solutions which exhibit blowup in finite time. In particular, the capillary pressure, Maxwell pressure and inertial pressure (i.e. kinetic energy per unit volume) undergo divergence near the liquid tip as $\tau^{-2/3}$ where $\tau \to 0$ denotes the blowup time. This divergent behavior is caused by the rapidly shrinking radius of curvature of the liquid tip which undergoes continuous field self-enhancement, subsequent to which ion emission is known to occur.  The numerically computed self-similar solutions describe a fully \textit{dynamic} conic tip that can adopt various internal flow configurations, representing rapidly accelerating or decelerating regions near the liquid apex which move in unison or in opposition to the bulk flow further from the tip.

The asymptotic self-similar solutions for the velocity potential, electric potential and interface shape properly describe their  functional behavior at distances far from the liquid apex. The near field behavior is not amenable to analytic solution and requires a numerical approach. The boundary integral technique developed for this purpose smoothly patches the behavior in the near field region of the conic apex to the functional behavior set by the far field  self-similar expansions. Unlike conventional boundary integral calculations on semi-infinite domains in which boundary contributions often rapidly decay and can be neglected, this patching technique accurately incorporates non-negligible boundary contributions along the far-field perimeter of the truncated liquid and vacuum domains.

The results of these numerical simulations highlight the crucial influence of inertial forces in the apical region, which are key to the unmasking of novel flow configurations we coin sub-conic, super-conic and mixed-conic. These refer to accelerating or decelerating flows whose liquid interface lies wholly below, wholly above or partially below and above the envelope set by the far field  solutions. Different choices of parameters for the family of solutions described yield flow configurations in which the liquid apical region can move in unison or in opposition to the the interior bulk flow. Since the inviscid approximation confers on the hydrodynamic system the property of time reversal symmetry, these simulations reveal both pre- and post-singularity behavior exhibiting features such as snapback from inertial recoil, tip bulging from rapid acceleration and other interesting dynamic behavior. The examples shown in Fig. \ref{fg:numresults} and Fig. \ref {fg:compareBT} also demonstrate that the local interior half-angle near the apex can be larger or smaller than the classic Taylor angle $\theta_T$. The fact that the apical angles differ from  $\theta_T$ was also observed in Fig. 3 of Ref. [\onlinecite{Albertson19}]. Those simulations \cite{FN3}, however, were conducted using moving mesh, finite element simulations of the full Navier-Stokes equation, while the approach adopted in this current study relies on solution of the inviscid interface Bernoulli equation.

The results revealed by our study vastly expands the spectrum of possible conic configurations associated with dynamic Taylor cone formation. Inertial forces play a critical role in shaping the accelerating or decelerating liquid front. Certain parameter choices for the family of asymptotic self-similar solutions described have also revealed the possibility of strong counterflow within the sharpening liquid tip accompanied by multiple stagnation points distributed along the moving interface. It remains to be seen whether these or other features described can give rise to interfacial instabilities, although we note that in this current study, numerical convergence for all configurations was always rapid and stable. We hypothesize that the self-similar Taylor conic envelope in the far field may be acting as a strong attractor which represses formation of any linear instabilities despite the enormous imbalance of surface forces acting on the system.

\acknowledgments
The authors gratefully acknowledge Dr. Peter Thompson for designing and maintaining the computing cluster used in this work. SMT also acknowledges valuable discussions with Dr. Colleen Marrese-Reading and Dr. James E. Polk at the NASA Jet Propulsion Laboratory involving  their ongoing experiments with liquid metal ion sources.


\begin{thebibliography}{44}
\bibitem [{FN1()}]{FN1}%
  \BibitemOpen
  \href@noop {} {}\bibinfo {note} {For reasons unknown, much of the elegant
  theoretical work that has been carried out by N. Zubarev and colleagues at
  the Institute for Electrophysics on the behavior of electrified liquids is
  often unreferenced. Those interested in the field are encouraged to review
  their large collection of insighful papers.}\BibitemShut {Stop}%
\bibitem [{FN2()}]{FN2}%
  \BibitemOpen
  \href@noop {} {}\bibinfo {note} {We note that our labels for the series
  coefficients $\{a_n, b_n\}$ refer to the expansions for the field variables
  $\psi$ and $\phi$, respectively. The labels in Ref. [\onlinecite{Zubarev01}]
  and Ref. [\onlinecite{Suvorov04}] are reversed.}\BibitemShut {Stop}%
\end{thebibliography}

%

\end{document}